\documentclass{article}
\usepackage{color}
\usepackage{graphicx}
\usepackage{amsmath, amsthm}
\usepackage{mathtools}
\usepackage{url}
\usepackage{soulutf8}
\RequirePackage[colorlinks,citecolor=blue,urlcolor=blue]{hyperref}
\usepackage{xcolor}
\usepackage{chngcntr}
\usepackage{subfig}
\usepackage{subcaption} 
\usepackage{booktabs}
\usepackage{tikz}
\usepackage{makecell}
\usepackage{hyperref}
\usepackage{algorithm}
\usepackage[noend]{algpseudocode}
\usepackage{setspace}

\usepackage{xfrac}
\definecolor{forest}{rgb}{0.133,0.545,0.133}
\usepackage{multirow}
\usepackage{amsfonts}

\evensidemargin=\oddsidemargin
\usepackage{etoolbox}

\newif\ifabbreviation
\pretocmd{\thebibliography}{\abbreviationfalse}{}{}
\AtBeginDocument{\abbreviationtrue}

\begin{document}
	\newcommand{\bb}{\boldsymbol{\beta}}

	\title{Joint estimation of the basic reproduction number and serial interval using Sequential Bayes}


	\author{Tatiana Krikella\footnote{Tatiana Krikella is the corresponding author and may be contacted at \url{krikella@yorku.ca}.} \hspace{35pt} Jane M. Heffernan \hspace{35pt} Hanna Jankowski \bigskip \\  
    \textit{Department of Mathematics \& Statistics} \\ \textit{York University, Toronto, ON, Canada, M3J 1P3}}
    
	\date{}

	\maketitle

	\begin{abstract}

Early in an infectious disease outbreak, timely and accurate estimation of the basic reproduction number ($R_0$) and the serial interval (SI) is critical for understanding transmission dynamics and informing public health responses. While many methods estimate these quantities separately, and a small number jointly estimate them from incidence data, existing joint approaches are largely likelihood-based and do not fully exploit prior information. We propose a novel Bayesian framework for the joint estimation of $R_0$ and the serial interval using only case count data, implemented through a sequential Bayes approach. Our method assumes an SIR model and employs a mildly informative joint prior constructed by linking log-Gamma marginal distributions for $R_0$ and the SI via a Gaussian copula, explicitly accounting for their dependence. The prior is updated sequentially as new incidence data become available, allowing for real-time inference. We assess the performance of the proposed estimator through extensive simulation studies under correct model specification as well as under model misspecification, including when the true data come from an SEIR or SEAIR model, and under varying degrees of prior misspecification. Comparisons with the widely used White and Pagano likelihood-based joint estimator show that our approach yields substantially more precise and stable estimates of $R_0$, with comparable or improved bias, particularly in the early stages of an outbreak. Estimation of the SI is more sensitive to prior misspecification; however, when prior information is reasonably accurate, our method provides reliable SI estimates and remains more stable than the competing approach. We illustrate the practical utility of the proposed method using Canadian COVID-19 incidence data at both national and provincial levels.

\end{abstract}

		\bigskip


	\maketitle

	\baselineskip=19.5pt


\section{Introduction}

At the beginning of a disease outbreak, the estimation of certain parameters is vital so that public health officials have an understanding of the disease in real time. Two such parameters are the basic reproduction number ($R_0$), and the generation interval. $R_0$ is the average number of secondary cases a primary case will infect, assuming a completely susceptible population \cite{boonpatcharanon2022estimating}, while the generation interval is the time between the infection of a primary case and one of its secondary cases \cite{svensson2007note}. Since the generation interval is difficult to estimate, and sometimes unobservable, the serial interval (SI) is often used as the parameter of interest, which is the time between the onset of symptoms in the primary case and the onset of symptoms in one of its secondary cases. The distributions of the generation time and SI have the same mean, provided that the times between infection and onset of symptoms are independent and identically distributed, but their variances may differ \cite{svensson2007note}. The $R_0$ and SI, together, provide important insights into an emerging infectious disease at the beginning of an outbreak. Accurate early estimates of these parameters are critical for informing public health interventions, modeling disease spread, and projecting healthcare needs \cite{britton2019estimation}. Misestimation, particularly of the serial interval, can bias reproduction number estimates and compromise real-time decision-making \cite{park2021forward}.

There have been many different ways of estimating the SI in the literature, however a common theme is to first assume that the serial interval follows either a normal, lognormal, gamma or Weibull distribution \cite{cori2013new, wallinga2004different, zhao2020estimating, talmoudi2020estimating, griffin2020rapid, cowling2009estimation, forsberg2008likelihood, nishiura2020serial, thompson2019improved, ganyani2020estimating, donnelly2011serial, becker2010type}. Once this distinction is made, some researchers have implemented frequentist methods in the form of maximumn likelihood estimation techniques \cite{zhao2020estimating, talmoudi2020estimating, cowling2009estimation, donnelly2011serial}, while others adopt a Bayesian approach \cite{thompson2019improved, ganyani2020estimating, becker2010type}.

A smaller set of studies considers the joint estimation of $R_0$ and the SI. One of the most well-known is the likelihood-based method of White and Pagano \cite{forsberg2008likelihood}. In this method, information on the number of cases observed each day is used to simultaneously estimate both $R_0$ and SI. This type of data is much more readily available at the start of an outbreak compared to contact tracing data. It is assumed that the number of secondary cases produced by an infected individual follows a Poisson distribution with mean $R_0$, and that the serial interval follows a multinomial distribution. To estimate, when the serial interval is unknown, the probabilities of the multinomial distribution are modelled as a two-parameter gamma distribution which is discretized, then to estimate $R_0$, $\alpha$, and $\beta$, the latter two parameters being those belonging to the gamma distribution, a maximum likelihood method is applied. 

An extension to the White and Pagano likelihood-based method was introduced where Bayesian methods were applied to allow for the inclusion of additional data as prior information via prior distributions \cite{moser2015impact}. This additional data can be, for example, contact tracing data. Then, to simultaneously estimate $R_0$ and SI, a Markov chain Monte Carlo (MCMC) method is used.

One thing to note across these various methods is that, often times, censoring or truncation is used when estimation the SI. For instance, in the White and Pagano method \cite{forsberg2008likelihood}, based on prior knowledge, a value $k$ may be defined such that the SI is known to be less than that value. Hence, making the distribution of the SI 0 when time exceeds $k$. In terms of truncation, this is sometimes applied, specifically left truncation, to account for instances when the primary case exhibits symptoms after their secondary case, resulting in a negative value of the SI \cite{cowling2009estimation}. 

To our knowledge, there have not been any methods proposed which use a Bayesian framework to jointly estimate $R_0$ and the SI solely from incidence data. A promising candidate is the sequential Bayes framework, in which a prior on the parameters of interest is updated iteratively as new case data becomes available. This framework has been applied to estimate $R_0$ using a non-informative (uniform) prior \cite{bettencourt2008real}. In many applications of Bayesian methods for $R_0$, priors tend to be non-informative, meaning they do not incorporate information from previous variants of the same disease or from similar outbreaks in other populations \cite{yang2013bayesian, narula2015bayesian, guo2023estimating}. In some examples, a very weakly informative prior is used, again to allow the data to drive the estimates \cite{moser2015impact, elderd2013population}. While $R_0$ can vary by population and setting, previous outbreaks of the same pathogen often provide valuable prior information, particularly when early data are limited. The SI, though more stable across settings, still benefits from joint modeling to account for uncertainty.

In this paper, we propose a joint Bayesian method to estimate both $R_0$ and the SI from case-count data using the sequential Bayes framework. Crucially, we introduce a mildly informative prior on $R_0$ and the SI instead of a flat prior, which can be informed by data from earlier variants or other populations.

The remainder of this paper is structured as follows. Section~\ref{s: methods} describes the sequential Bayes method and is assumptions, along with our prior formulation. Section~\ref{s: simStudies} is a comprehensive set of simulation studies which demonstrate the behaviour of our proposed estimator, including how robust the estimates are under mispecification. We test our method using real Canadian COVID-19 data in Section~\ref{s: realExample}, where, specifically, we use data from the whole country, as well as the three most populous provinces.

\section{Methods}
\label{s: methods}

\subsection{SIR Model}
\label{ss: SIR}

Assume we have case count data that comes from an \textit{SIR model}, which is a system of three ordinary differential equations (ODEs):

\begin{align*} 
\frac{dS}{dt} &=  -\beta \frac{S(t)}{N}I(t) \\ 
\frac{dI}{dt} &=  \beta \frac{S(t)}{N}I(t)-\gamma I(t) \\ 
\frac{dR}{dt} &= \gamma I(t)
\end{align*}

Here, $S(t)$ consists of all individuals susceptible to the infection, $I(t)$ is all of the infected individuals, and $R(t)$ are the recovered individuals. The parameter $\beta$ is the transmission rate, which is the average number of people a disease carrier infects per day. The parameter $\gamma$ is the rate at which infected people recover. Note that $1/\gamma$ is the average number of days that someone stays contagious, and is equivalent to the serial interval (SI). Finally, $N = S(t) + I(t) + R(t)$.

In an SIR model, $\frac{\beta}{\gamma} = R_0$. When $\beta > \gamma$, then $R_0 > 1$ and the disease will spread. When $\beta < \gamma$, then $R_0 < 1$ and the disease dies out. We emphasize the fact that $\gamma$ and $R_0$ are not independent.

\subsection{Sequential Bayes Estimation}
\label{ss: seqBayes}

The idea behind the Sequential Bayes method is to assume a prior on the parameter of interest, then update the prior sequentially \cite{bettencourt2008real}. At each new time point, the new prior is the posterior at the previous time point. This method is based on case count data, and it is assumed that these infectious counts are observed at periodic times (i.e. days or weeks). This approach also assumes that the data comes from an SIR model.

Under the SIR model, and considering a time interval $t_{j+1} - t_j$, we have 

\begin{equation} 
\begin{split}
I(t_{j+1}) & = I(t_j)\exp\left[{\gamma \int_{t_j}^{t_{j+1}}\left(R_0 \frac{S(s)}{N} -1\right)ds}\right] \\
 & \approx I(t_j)\exp\left[(t_{j+1} - t_j)\gamma(R_t-1)\right]
\end{split}
\label{eq: infections}
\end{equation}

where $R_t \approx R_0$ at the beginning of an infection. 

Using the result from Eq~\eqref{eq: infections}, we assume that the conditional distribution of $I(t_{j+1})$, conditional on $I(t_j)$, $R_0$, and $\gamma$ is Poisson with mean $\lambda = I(t_j)\exp\left[(t_{j+1} - t_j)\gamma(R_t-1)\right]$. In our work, we are interested in estimating both $\gamma$ and $R_0$ using a joint prior. 

\subsection{Prior Definition} 
\label{ss: priorDef}

Since $R_0$ and $\gamma$ are dependent, we define a marginal distribution for each parameter of interest and join them via a copula to construct a joint distribution. Specifically, we use a log-Gamma distribution as the marginal distribution for both $R_0$ and $\gamma$ as this is the conjugate prior of $\theta = (t_{j+1} - t_j)\gamma(R_0-1)$ (see Supplementary Material, Section S1). If we define $\theta$ this way, then we can write $I(t_{j+1})|I(t_j), R_0, \gamma \sim Poisson(I(t_j)\exp[\theta]).$

In general, if $X \sim Gamma(\alpha, \beta)$, then $Y = \ln X$ follows a log-Gamma distribution with shape and scale parameters, $\alpha$ and $\beta$, respectively. The probability density function of the log-Gamma distribution is shown in Eq~\eqref{eq: logGamma}. 

\begin{equation}
f(y) = \frac{1}{\Gamma(\alpha)}\left[\frac{e^y}{\beta}\right]^\alpha e^{-e^y/\beta}
 \label{eq: logGamma}
\end{equation}

The mean and variance of a log-Gamma distributed random variable, $Y$, are shown in Eqs~\eqref{eq: mean} and \eqref{eq: var}, respectively, where $\psi(\alpha)$ is the digamma function evaluated at $\alpha$ and $\psi^{'}(\alpha)$ is the trigamma function evaluated at $\alpha$. The digamma function is the logarithmic derivative of the gamma function, while the trigamma function is the derivative of the digamma function.

\begin{equation}
    E(Y) = \psi(\alpha) + \ln \beta 
    \label{eq: mean}
\end{equation}

\begin{equation}
    Var(Y) = \psi^{'}(\alpha)
    \label{eq: var}
\end{equation}

To define the joint prior, we join the two log-Gamma marginal distributions of $R_0$ and $\gamma$ with a Gaussian copula which models the dependence between the two parameters. The joint prior for $R_0$ and $\gamma$ is shown in Eq~\eqref{eq: jointPrior}, where $F_{R_0}(R_0)$ and $F_{\gamma}(\gamma)$ represent the cumulative distribution functions of $R_0$ and $\gamma$, respectively, $f_{R_0}(R_0)$ and $f_{\gamma}(\gamma)$ represent the log-Gamma marginal distributions of $R_0$ and $\gamma$, respectively, and the Gaussian copula ($c_\rho $) is applied using correlation parameter, $\rho \in [-1,1]$. Since $R_0 = \beta/\gamma$, $\rho$ will be negative.

\begin{equation}
    \pi (R_0, \gamma) = c_\rho (F_{R_0}(R_0)F_{\gamma}(\gamma))f_{R_0}(R_0)f_{\gamma}(\gamma)
    \label{eq: jointPrior}
\end{equation}

\subsection{Note on Relationship between $R_0$ and $\gamma$}
\label{ss: theoreticalCurve}

At the beginning of an outbreak, when most individuals are still susceptible, the number of new infections tends to grow exponentially \cite{fisman2013idea}. This growth can be approximated using the relationship shown in Eq.~\eqref{eq: exp-growth}. 

\begin{equation}
\label{eq: exp-growth} 
    I(t) = {R_0}^{t\gamma}
\end{equation} 

This formulation reflects the idea that the number of infections multiplies by a factor of $R_0$ over successive serial intervals. Since each infectious period lasts on average $1/\gamma$ units of time (i.e., the serial interval), the term $t\gamma$ represents the number of serial intervals that have occurred by time $t$. It is important to ensure that the units of time used for $t$ and $\gamma$ are consistent (i.e., if $t$ is in weeks, ensure that $\gamma$ is also represented in weeks). 

We highlight this relationship because we will use it as a theoretical reference curve to assess whether our posteriors are updating in a way that reflects expected early incidence case counts. It is important to note, however, that this theoretical curve is a rough, large-scale approximation of epidemic development. One issue that could arise, for example, is if $1/\gamma$ is large, this corresponds to a very large $R_0$, which may indicate that the model is not suitable for describing the epidemic. Further, this model is only valid during the early weeks of an outbreak. As the epidemic progresses, the assumptions underlying $R_0$, specifically, that everyone in the population is susceptible, are increasingly violated, and the curve, which is a function of $R_0$, may no longer be a good representation of the dynamics.

\section{Simulation Studies}
\label{s: simStudies}

We conduct simulation studies to assess the performance of our proposed sequential Bayes method for jointly estimating $R_0$ and the SI (equivalently, $1/\gamma$). We compare its performance to that of the White and Pagano likelihood-based method of joint estimation \cite{forsberg2008likelihood}.

For each simulation scenario, we generate weekly case count data under one of three epidemic models: SIR, SEIR, or SEAIR. The purpose of including SEIR and SEAIR simulations is to evaluate the robustness of our method under model misspecification, since our inference assumes the data arises from an SIR model. Details of the SEIR and SEAIR models can be found in Supplementary Material Section S2. Each dataset consists of 1000 simulated epidemic trajectories (i.e., rows), with columns corresponding to weekly case counts. 

We consider two diseases: Influenza 1 and Influenza 2. For Influenza 1, the true value of $R_0$ is $5/3$ in the SIR and SEIR scenarios, and $7/3$ in the SEAIR scenario. For Influenza 2, $R_0$ is $5/3$ across all three models. The true SI for Influenza 1 is 5 days (i.e., $5/7$ weeks) for the SIR dataset, and 8 days (i.e., $8/7$ weeks) for the SEIR and SEAIR datasets. For Influenza 2, the true SI is 5 days for all datasets.

We define the support of $R_0$ to be $[0.001, \kappa]$, where $\kappa = 10$, and the support of $\gamma$ to be $[0.001, \eta]$, where $\eta = 5$. These bounds can be adjusted depending on prior knowledge about the disease.

We use the prior described in Section~\ref{ss: priorDef} for all simulation studies. The hyperparameters of the prior differ depending on the dataset and whether we are assessing performance under prior misspecification, however, for all cases, we fixed the hyperparameter $\alpha$ for both marginal distributions to be 2. This hyperparameter controls the variance of the marginal distributions, and by setting $\alpha = 2$, we ensure the prior is mildly informative. We outline the different cases used to test misspecification in Section~\ref{ss: simulationResults}.

\subsection{Illustrative example of posterior behavior}

\begin{figure}
    \centering
    \includegraphics[width=1\linewidth]{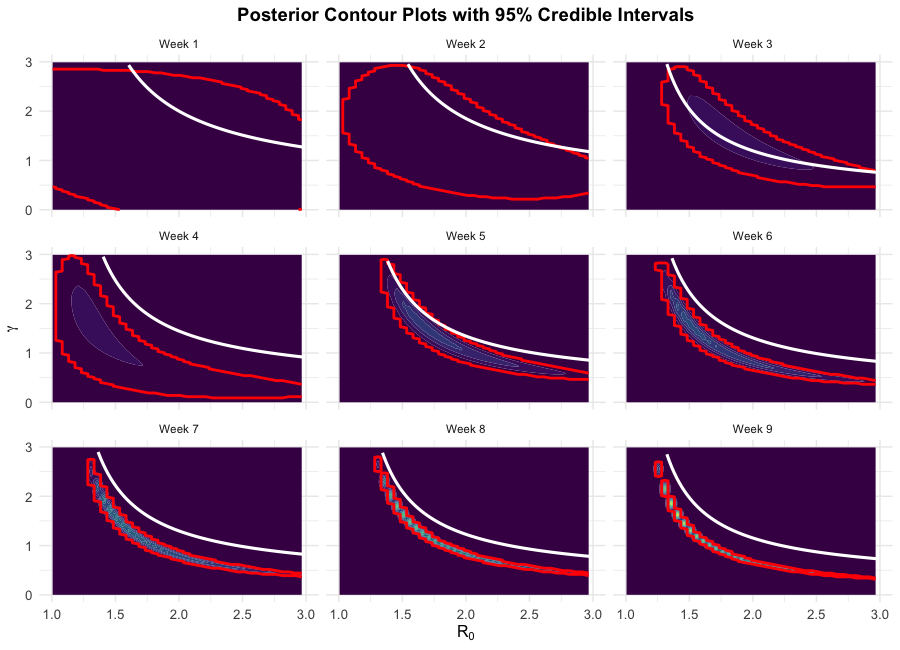}
    \caption{Each panel shows the prior that is used for that specific week represented by a contour plot, which, except for week 1, is the posterior from the previous week. The density is shown such that the more vibrant colours indicate higher density regions. The red lines denote 95\% credible intervals (CrI) for the prior used in each week, which were found using the highest density regions (HDR). Specifically, within each CrI is 95\% of the density, i.e. $P((\gamma, R_0) \in CrI) = 0.95$.}
    \label{fig: posteriorTracking}
\end{figure}

Before presenting aggregate results, we include a representative example to illustrate how the joint posterior behaves in a single simulation. Figure~\ref{fig: posteriorTracking} shows the posterior distribution from the first row of the Influenza 1–SIR dataset using a well-specified prior, i.e. the means of the marginal distributions used to construct the prior match the true values of the parameters. We show an example using a misspecified prior in Supplementary Material Section S3. In each panel of Figure~\ref{fig: posteriorTracking}, the prior that is used for the specific week is shown. In a sequential Bayes framework, this prior is simply the posterior from the previous week (except for Week 1, which is the original prior used). Overlaid is the theoretical curve $I(t) = {R_0}^{t\gamma}$, reflecting the expected growth in incidence during the early stage of an outbreak. As discussed in Section~\ref{ss: theoreticalCurve}, this curve provides a benchmark for identifying combinations of $R_0$ and $\gamma$ consistent with epidemic dynamics. We also include 95\% credible intervals (CrIs), shown as a red line, which are found using the highest density region (HDR) of the joint posterior. In early weeks, the posteriors are uniformly flat regions. However, as the sample size increases, i.e. as more weeks of data are added, the data takes over and the posterior begins to take the form of the theoretical curve. Further, we notice that as the sample size increases, the 95\% CrI becomes more precise compared to earlier weeks. The higher density can be seen within these regions, indicated by the brighter colours.

\subsection{Sensitivity to Prior Misspecification} 
\label{ss: priorMisspecification} 

To identify representative scenarios for evaluating the robustness of our method under model misspecification, we conducted an extensive simulation study across a discrete grid of parameter values using the Influenza 1-SIR dataset. For each scenario, we obtained posterior estimates of both $R_0$ and $SI = 1/\gamma$ at weeks 4, 5, and 6, and quantified estimation error using the L1 distance defined in Equation~\eqref{eq:L1}. We note that the units of $SI$ are in days. Weeks 4-6 were selected because preliminary analyses as they occur immediately prior to the inflection point of the data.

\begin{equation}
\label{eq:L1}
\mathrm{L1}
=
\left| \mathrm{median}_{4} - \theta_{\text{true}} \right|
\;+\;
\left| \mathrm{median}_{5} - \theta_{\text{true}} \right|
\;+\;
\left| \mathrm{median}_{6} - \theta_{\text{true}} \right|.
\end{equation}

We evaluated all combinations in an $11 \times 11$ grid (121 total cases). The values considered for $R_0$ are
\[
R_0 \in \left\{
1.00, 1.17, 1.33, 1.5, \mathbf{1.66}, 1.83, 2.00,
2.17, 2.33, 2.67, 3.00
\right\},
\]
and for $SI$ are
\[
SI \in \left\{
2.00, 3.00, 3.50, 4.00, 4.50, \mathbf{5.00}, 5.50, 6.00, 6.50, 7.00, 8.00
\right\}.
\]

To summarize performance across the grid, we constructed contour plots of the L1 differences, with the specific parameter pairs we tested superimposed. Two sets of contours were produced: one for the estimation of $R_0$ and one for the estimation of $SI$. In each plot, $R_0$ is displayed on the horizontal axis and $SI$ on the vertical axis. The axes intersect at the true parameter values used in all simulations ($R_0 = 5/3$ and $SI = 5$). 

\begin{figure}
    \centering
        \subfloat[$R_0$]{%
            \includegraphics[width=.453\linewidth]{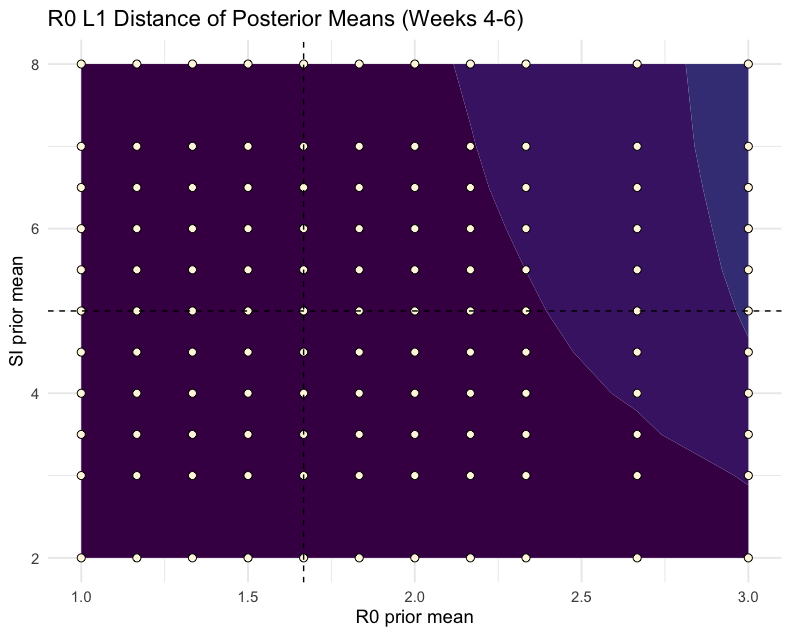}%
            
        } 
        \subfloat[$SI$]{%
            \includegraphics[width=.5\linewidth]{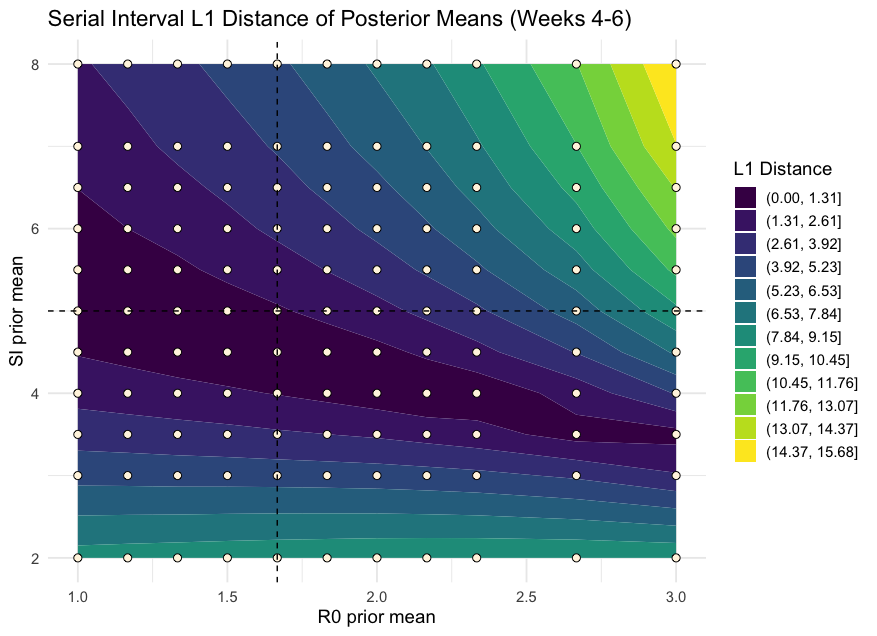}%
            
        } \\ 
        \caption{Plotting the L1 Differences of the medians of weeks 4, 5, and 6 under 121 cases.} 
        \label{fig: intensiveSimulation}
    \end{figure}

The results of this intensive simulation study are shown in Fig.~\ref{fig: intensiveSimulation}. Estimates of $SI$ are substantially more sensitive to misspecification than those of $R_0$, as demonstrated by noticeably larger L1 differences across the grid. Even under misspecification the estimates of $R_0$ have relatively low bias. The worst performance for $R_0$ and $SI$ occurs when both parameters are over-specified (e.g., $R_0 = 3$ and $SI = 8$). 

Based on these results, we selected five representative misspecified scenarios, along with the well-specified case, for subsequent simulations (Table~\ref{tab: misspecifiedCases}). For datasets in which the true parameter values differed from those in the Influenza 1–SIR dataset (i.e., Influenza 1–SEIR and Influenza 1–SEAIR), we applied equivalent shifts from the true values. That is, each misspecification was defined relative to the dataset-specific true $R_0$ and $SI$, preserving the same distance from truth as in the Influenza 1–SIR cases.

\begin{table}[h!]
\centering
\begin{tabular}{lcc}
\toprule
\textbf{Case} & $SI$ & $R_0$ \\
\midrule
Misspecification 1 & $4$ & $2$ \\
Misspecification 2 & $4$    & $1.33$      \\
Misspecification 3 & $6.5$   & $2.17$     \\
Misspecification 4 & $2$   & $3$ \\
Misspecification 5 & $8$   & $3$     \\
\bottomrule
\end{tabular}
\caption{Selected misspecified parameter scenarios used in subsequent simulations.}
\label{tab: misspecifiedCases}
\end{table}

Misspecifications~1 and 2 represent a moderate deviation from the truth: the serial interval is misestimated by 1 day and $R_0$ by $1/3$. Misspecification~3 assumes a larger estimate of the SI and $R_0$, however, in Figure~\ref{fig: intensiveSimulation}, this misspecification corresponds to an SI estimate with moderate bias and an $R_0$ estimate of very little bias. Misspecifications~4 and~4 correspond to severe misspecification of both parameters, with Misspecification~5 being the single worst-performing scenario in the grid. In this case, the serial interval is overestimated by 3~days and $R_0$ is overestimated by $4/3$.

\begin{figure}
    \centering
        \subfloat[Well-Specified]{%
            \includegraphics[width=.33\linewidth]{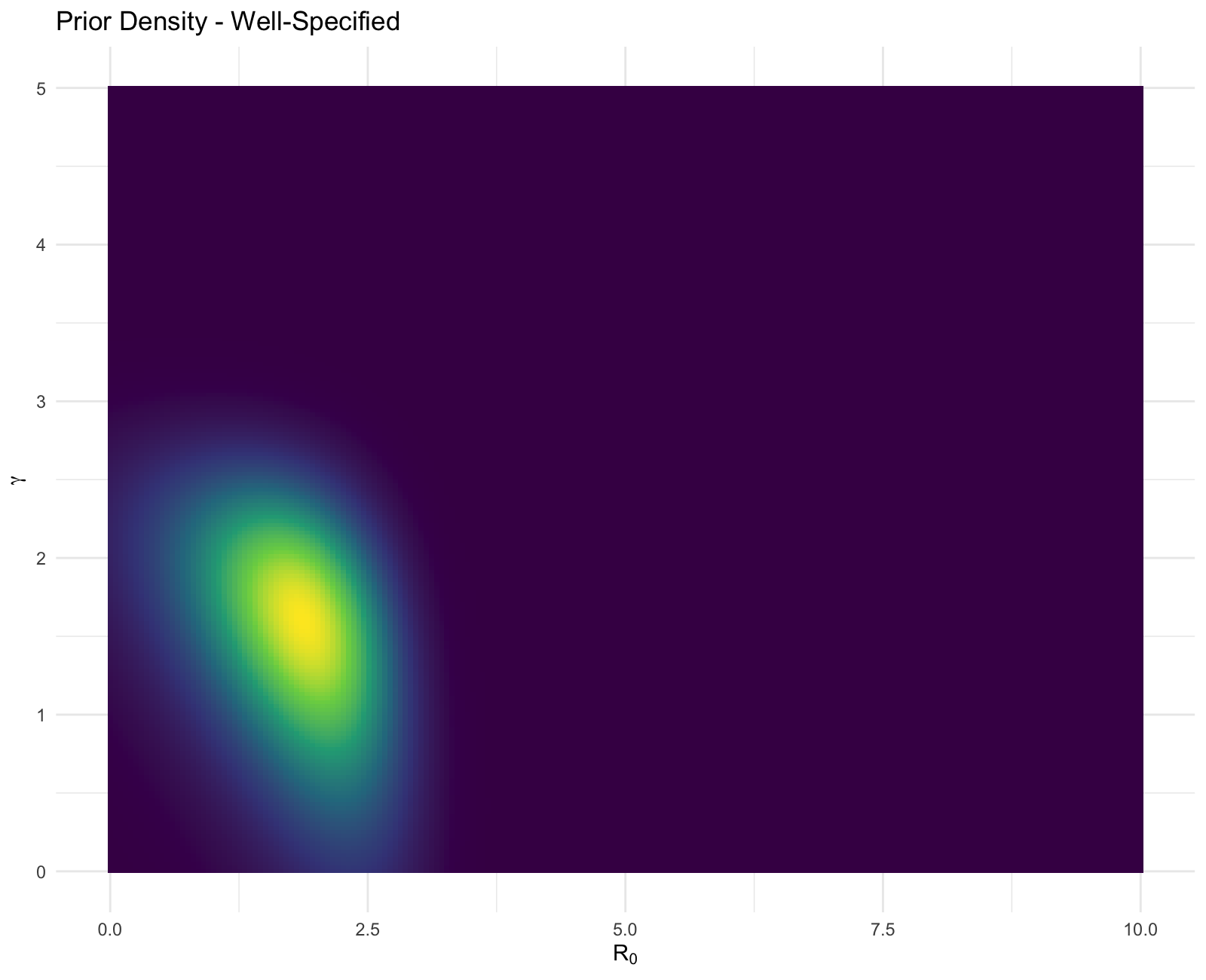}%
            
        } 
        \subfloat[Misspecification 1]{%
            \includegraphics[width=.33\linewidth]{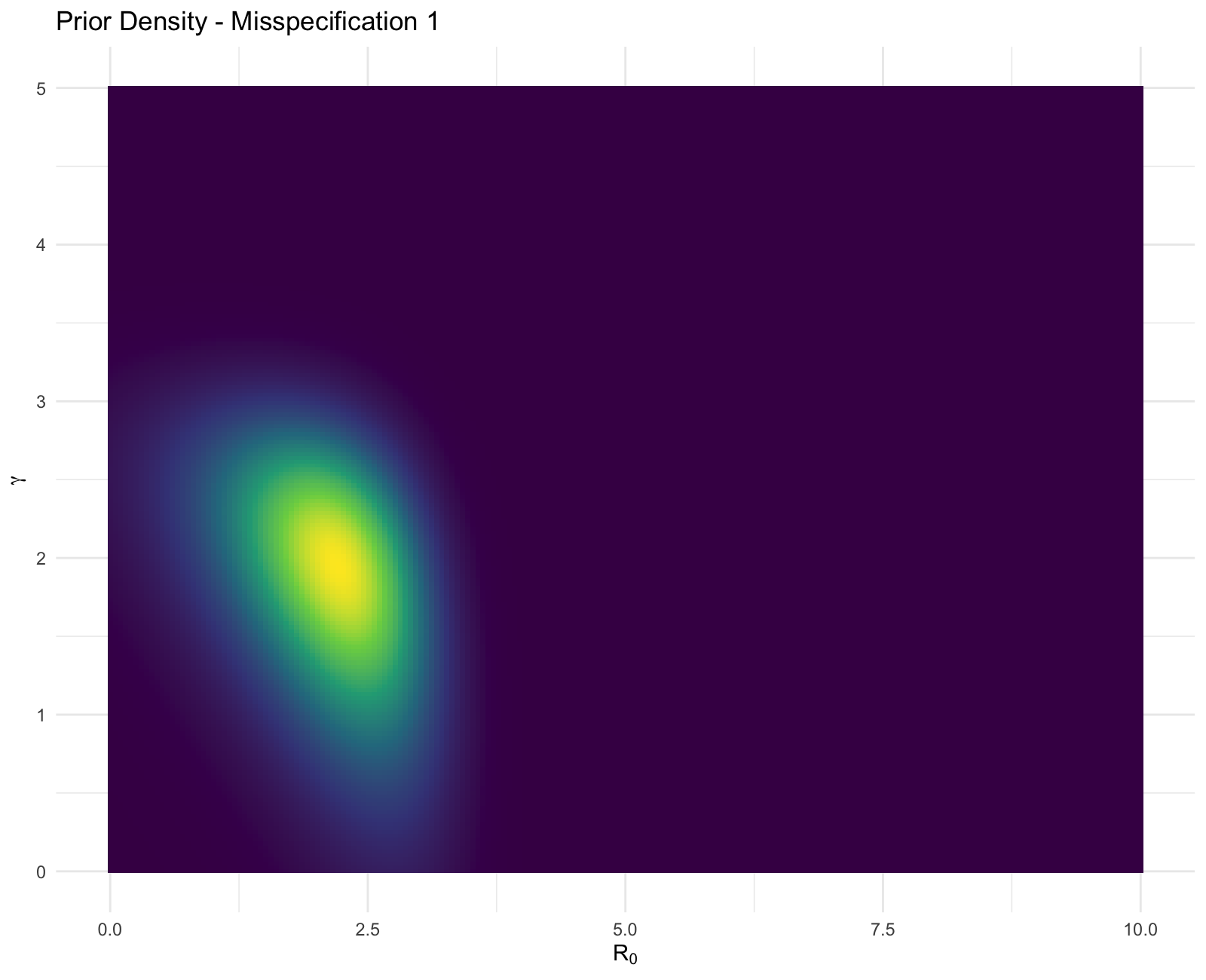}%
            
        } 
        \subfloat[Misspecification 2]{%
            \includegraphics[width=.33\linewidth]{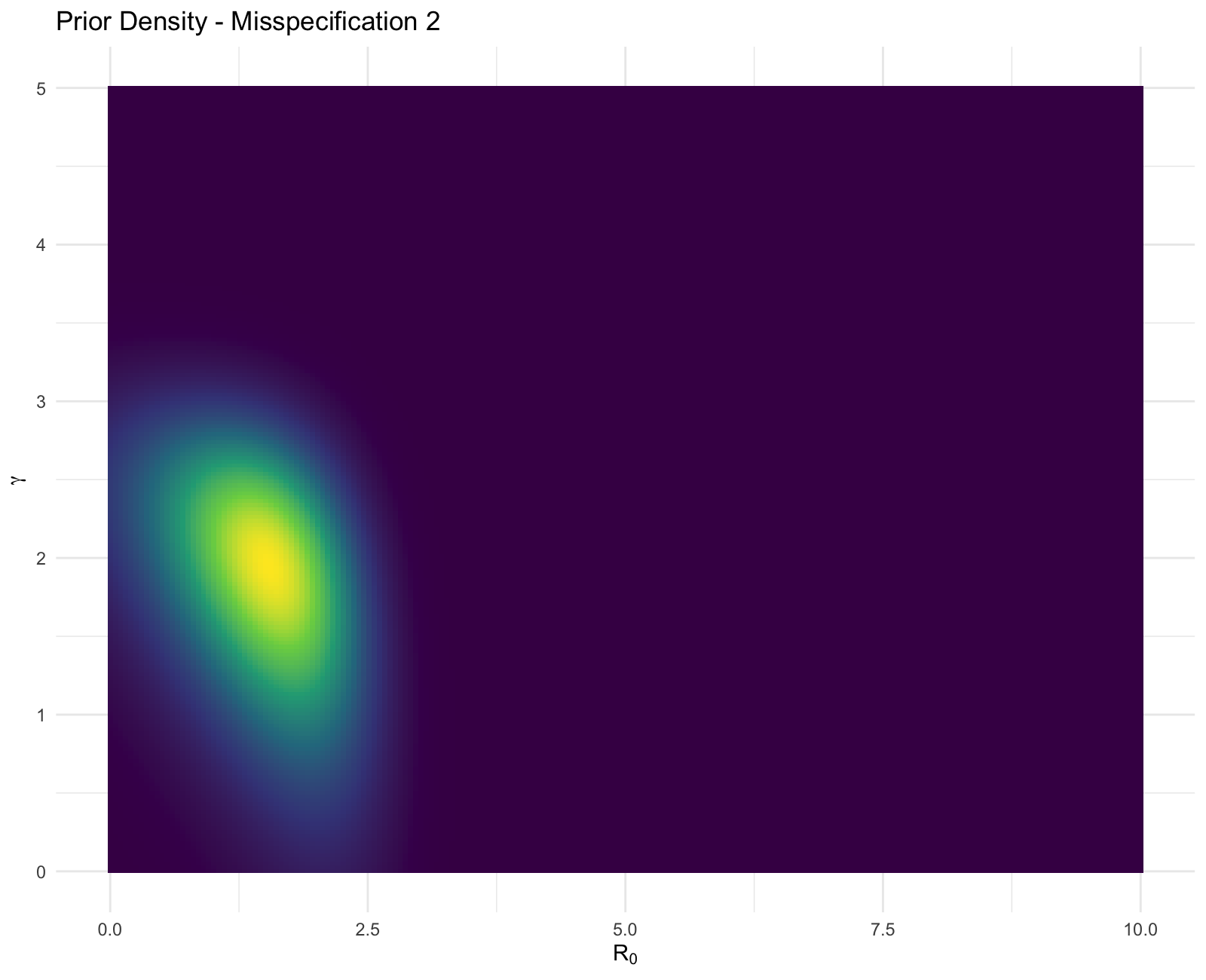}%
            
        } \\ 
        \vspace{1em}
        \subfloat[Misspecification 3]{%
            \includegraphics[width=.33\linewidth]{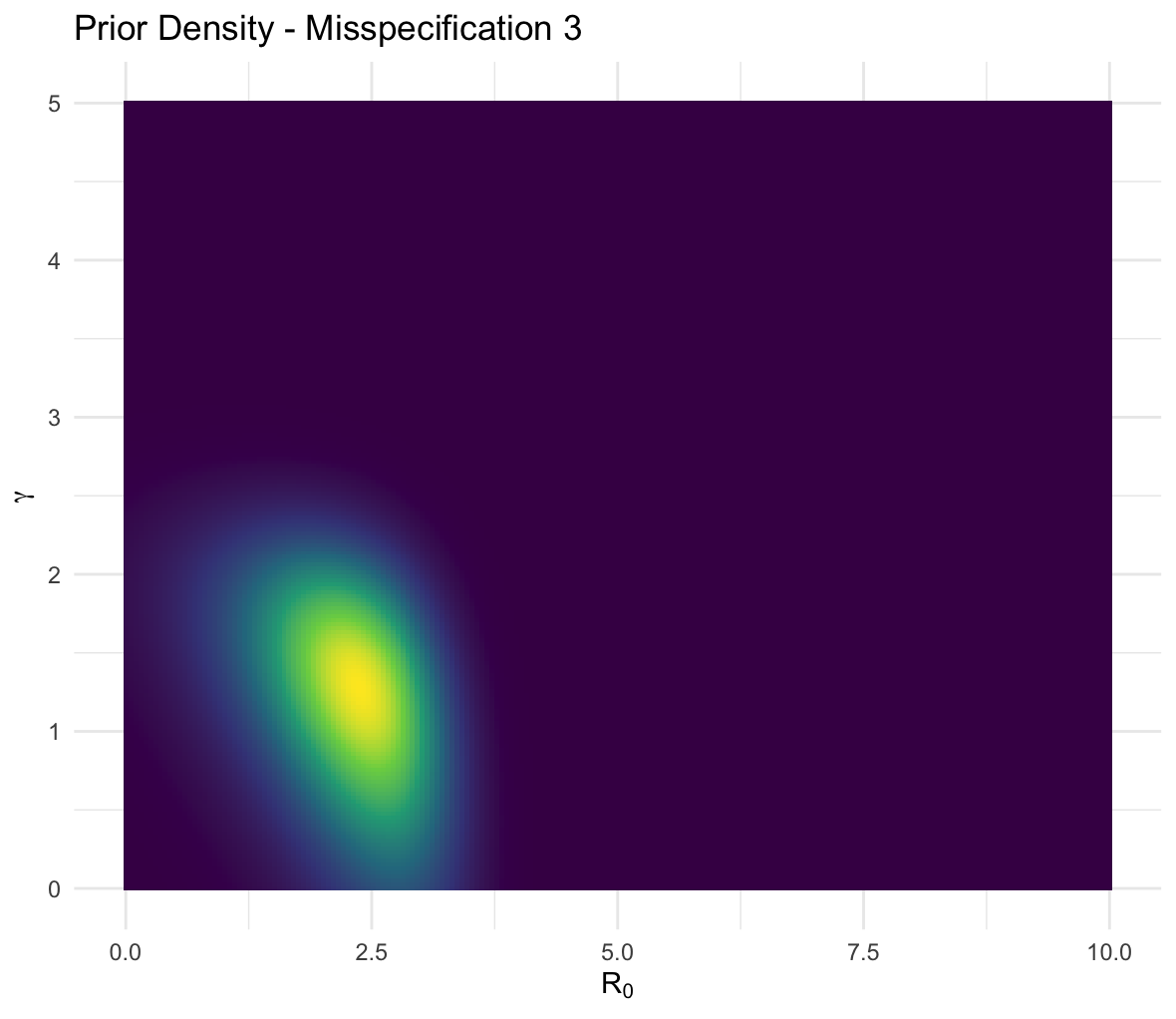}%
            
        } 
        \subfloat[Misspecification 4]{%
            \includegraphics[width=.33\linewidth]{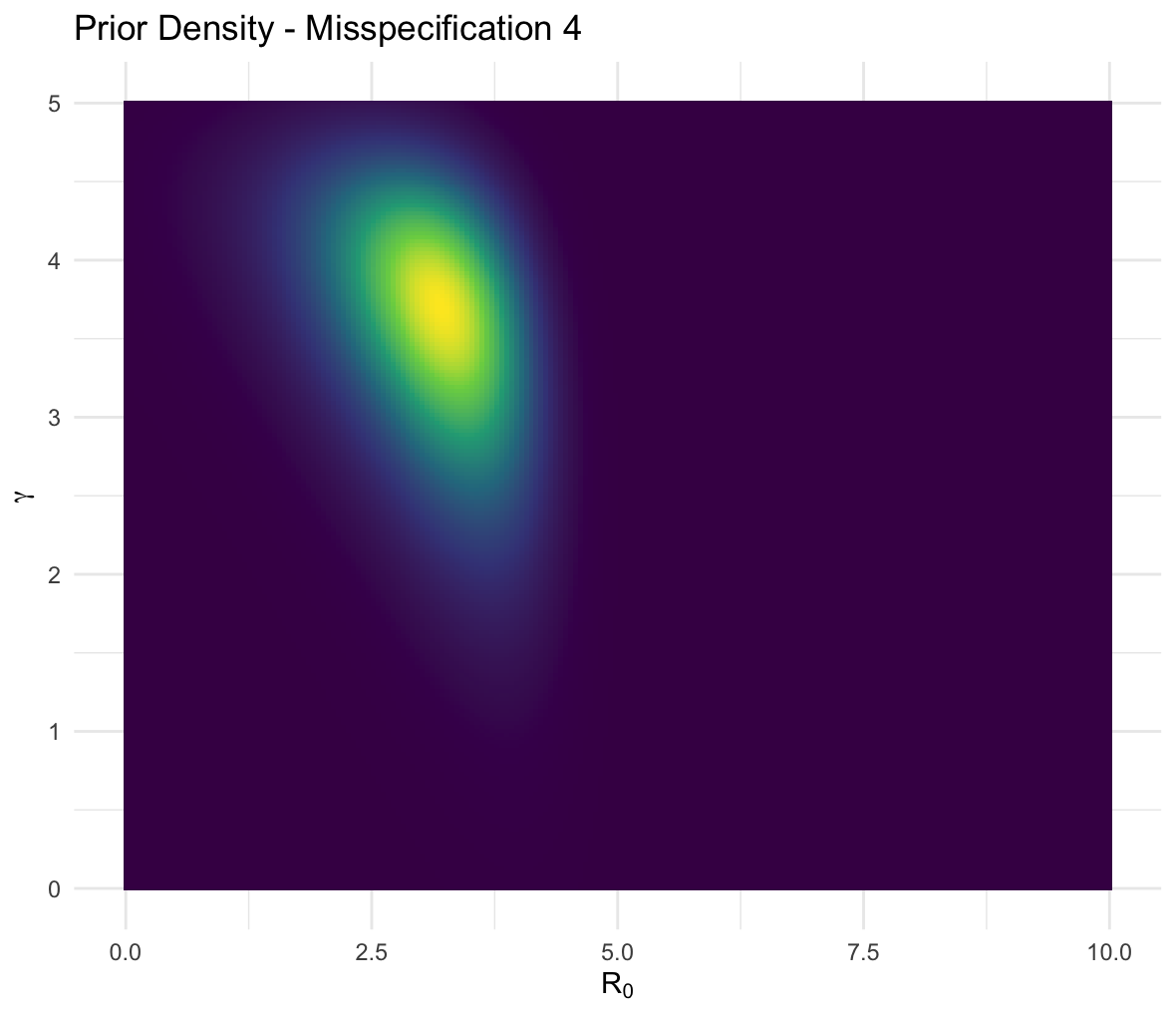}%
            
        } 
        \subfloat[Misspecification 5]{%
            \includegraphics[width=.33\linewidth]{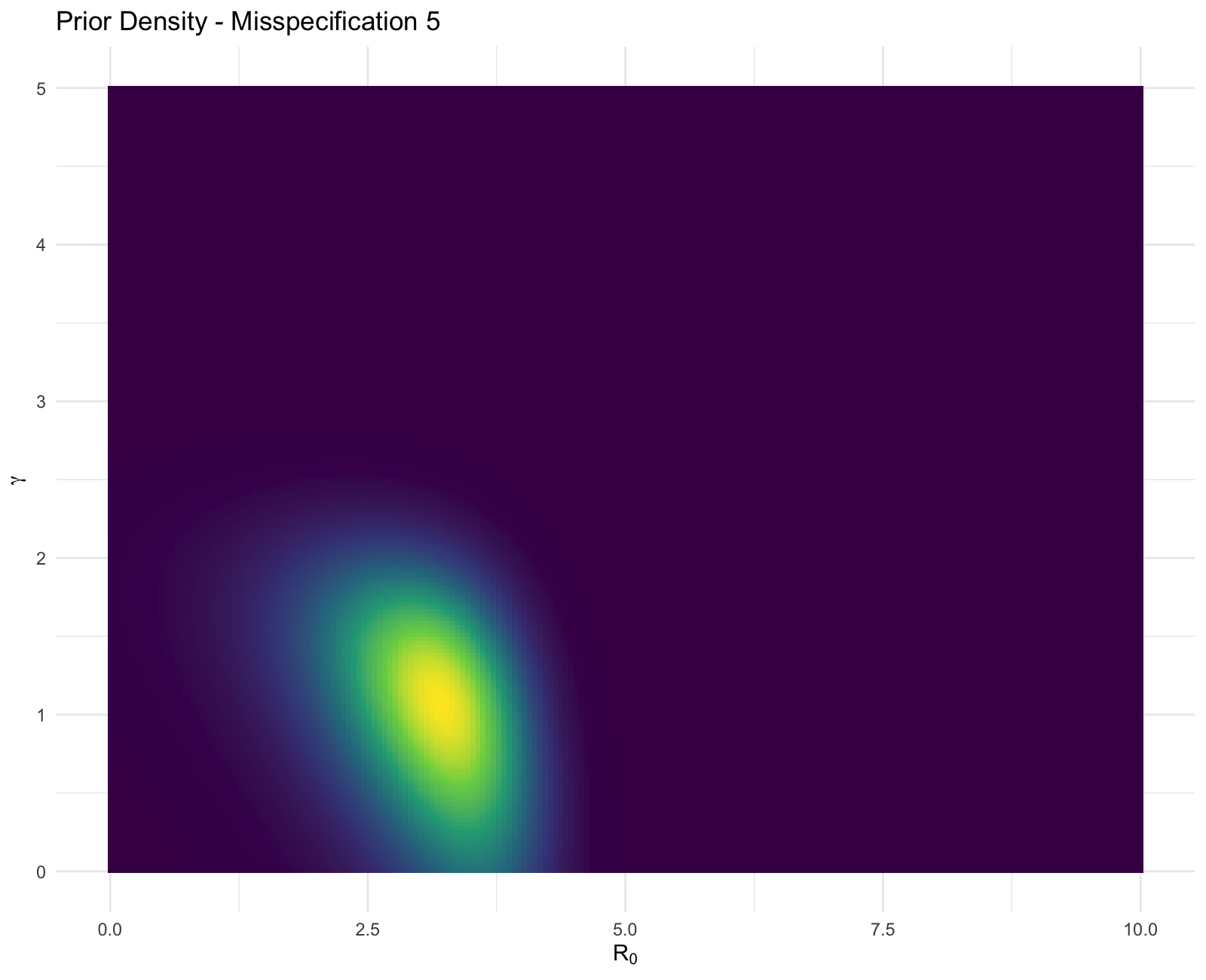}%
            
        } \\ 
        \caption{Joint prior density used in sequential Bayes for the well-specified, as well as the five misspecified, cases. The parameter $\gamma = 7/SI$, where SI is in days, is on the y-axis, and $R_0$ on the x-axis. The brighter colours indicate areas of higher density.} 
        \label{fig: priors}
    \end{figure}

We have included heat maps of each prior in Fig~\ref{fig: priors}. Note that, as outlined in Section~\ref{ss: priorDef}, the joint prior is placed on the parameters $R_0$ and $\gamma$, where $SI = 1/\gamma$. Thus, the heatmaps show the $\gamma$ on the y-axis instead of the SI. Further, since the data we are using is weekly case count data, although we list the SI in days for ease of interpretation in Table~\ref{tab: misspecifiedCases}, the prior needs to be in the same units as the data. Thus, $\gamma = 7/SI$, where SI is as listed in the aforementioned Table, to account for the units of the simulated data. The prior shown in Fig~\ref{fig: priors}(a) is equivalent to the first panel (i.e. Week 1 panel) in Fig~\ref{fig: posteriorTracking} where we display the updating of the posterior. 


\subsection{Estimation Performance} 
\label{ss: simulationResults}

We consider both Influenza 1 and 2 in this subsection and display the estimation performance of both the serial interval (SI) and $R_0$ in the form of boxplots. Each panel of the boxplot figures include the results from the White and Pagano method, the sequential Bayes method under a well-specified prior, and then five cases where the prior is misspecified as outlined in Table~\ref{tab: misspecifiedCases}. Included in these panels are the true value of the parameter of interest represented by a dotted red line, as well as a dotted blue line to indicate the inflection point of each dataset. We have included plots where the range of weeks only extends to the inflection point, but the full plots, including those weeks past the inflection point, can be found in Supplementary Material Section S4. 

Fig~\ref{R0_Influenza} shows the estimates of $R_0$ for the six datasets considered. Fig~\ref{R0_Influenza}(a) and (b) show the results under a correctly specified model, i.e. an SIR model. Immediately it is clear that the variance, even under misspecification, of our proposed sequential Bayes method is much smaller than that of the White and Pagano method across all weeks. Under both methods, the variance decreases as we have more data, but the sequential Bayes estimates are noticeably much more precise, especially in Weeks 5, 6, and 7. Moreover, it is in those same weeks where we see the smallest bias in the sequential Bayes estimates. Recall that Misspecification~5 was the worst performing scenario in the sensitivity analysis displayed in Fig~\ref{fig: intensiveSimulation}. This prior specification does not perform well compared to the other sequential Bayes estimates, however, when comparing to the White and Pagano estimate, has a much smaller variance and a better, or at least comparable, bias up until the inflection point of the dataset.

Fig~\ref{R0_Influenza}(c)-(f) display the results when using a misspecified model, where the data is simulated from an SEIR or SEAIR model. Similar to what was seen under the SIR model, the sequential Bayes estimates have a considerably smaller variance compared to the White and Pagano method, and the bias is much better, or at least similar, to the White and Pagano method up until the inflection point. The difference we see in these cases from the SIR case, specifically in Influenza 1, is that the two worst performing scenarios from the sensitivity analysis, instead of just the worst, resulted in a slightly higher bias compared to the other sequential Bayes estimates. In the SEIR case, this bias is higher than that of the White and Pagano estimates, but in the SEAIR case, the bias of the worst performing sequential Bayes cases is much better than the bias of the White and Pagano estimates.

\begin{figure}
    \centering
        \subfloat[Influenza 1, SIR]{%
            \includegraphics[width=.4\linewidth]{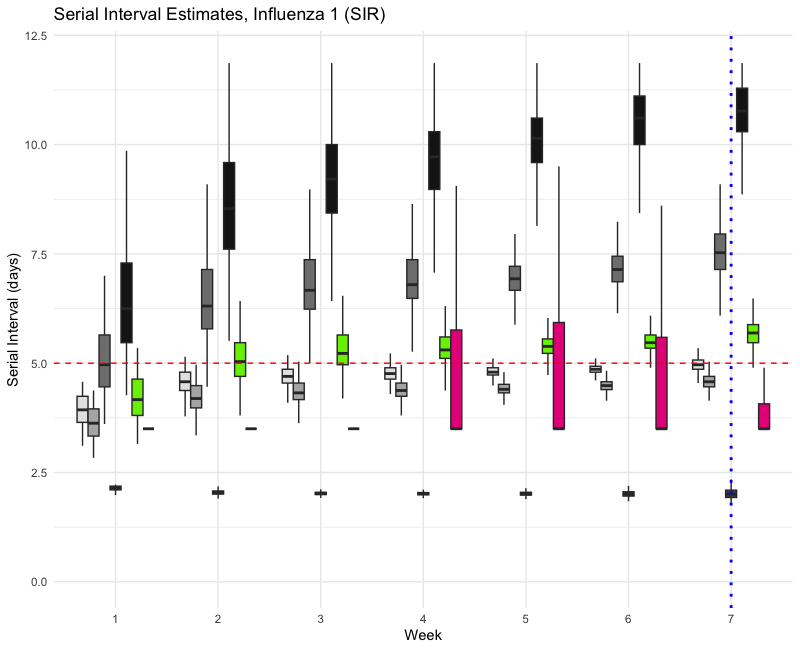}%
            
        } 
        \subfloat[Influenza 2, SIR]{%
            \includegraphics[width=.4\linewidth]{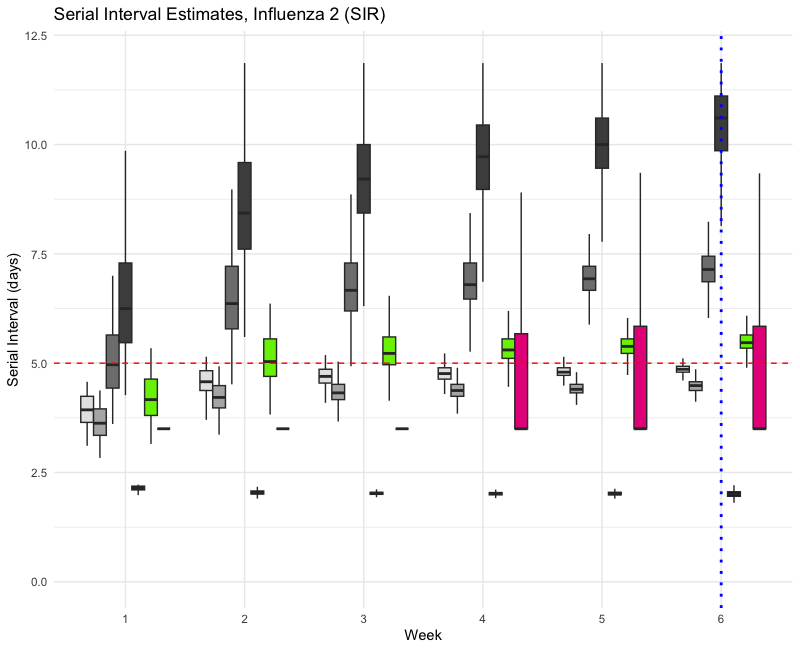}%
            
        } \\
        \vspace{1em} 
        \subfloat[Influenza 1, SEIR]{%
            \includegraphics[width=.4\linewidth]{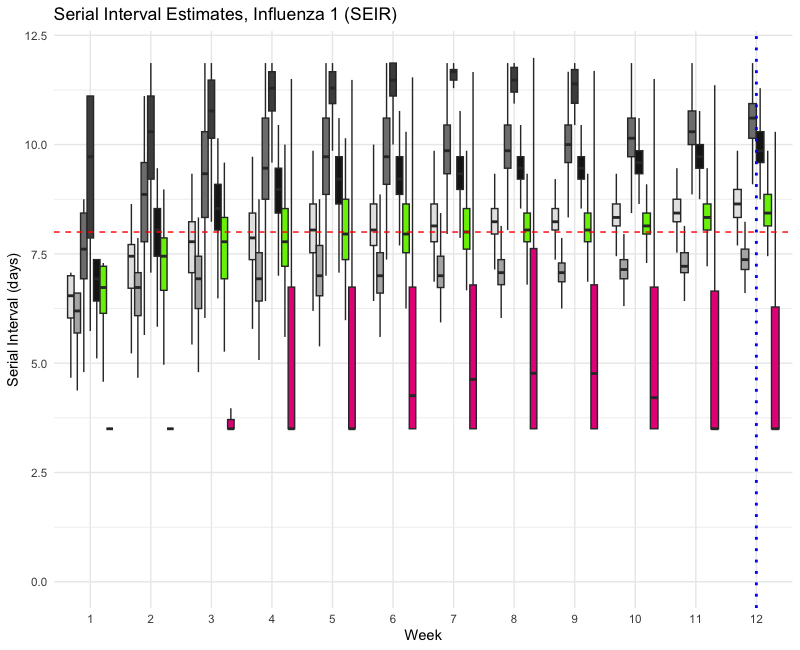}%
            
        }
        \subfloat[Influenza 2, SEIR]{%
            \includegraphics[width=.4\linewidth]{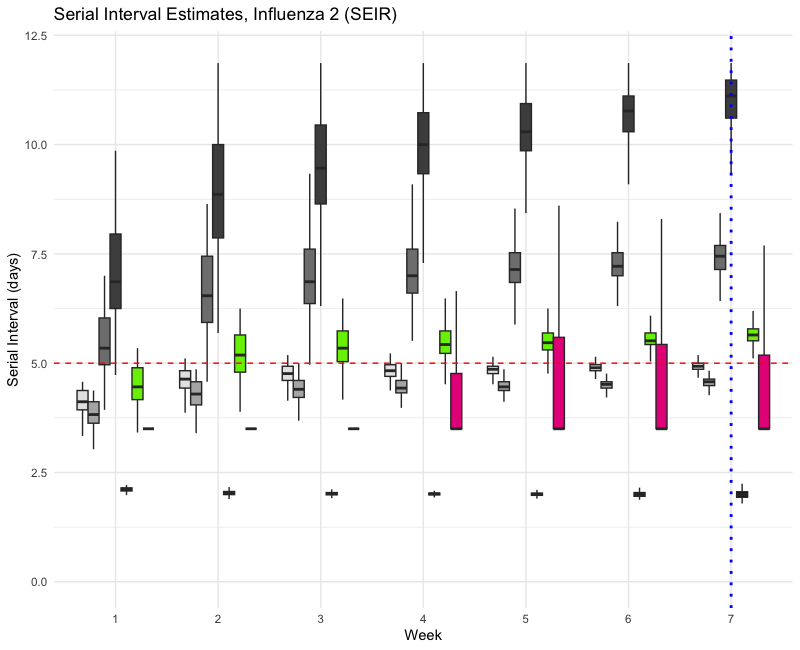}%
            
        }\\
        \vspace{1em} 
        \subfloat[Influenza 1, SEAIR]{%
            \includegraphics[width=.4\linewidth]{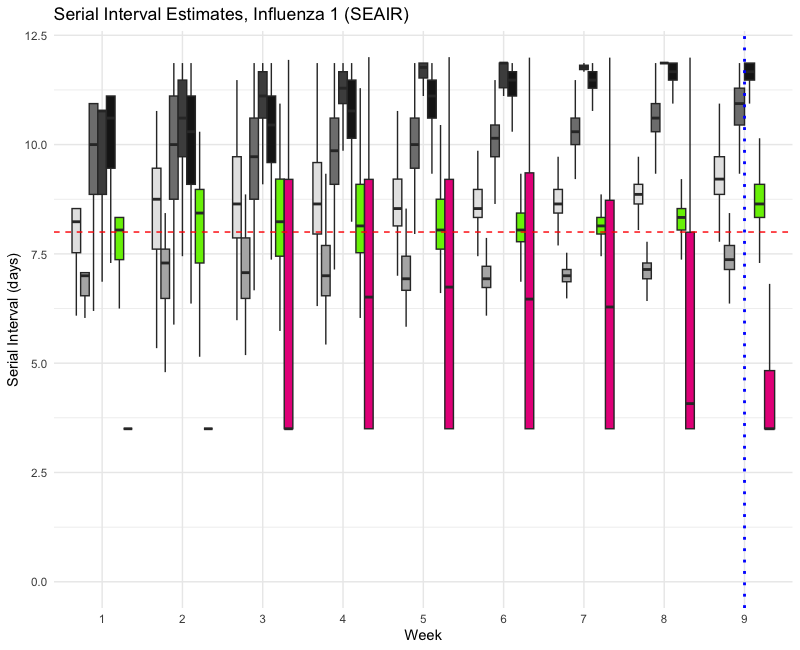}%
            
        }
        \subfloat[Influenza 2, SEAIR]{%
            \includegraphics[width=.4\linewidth]{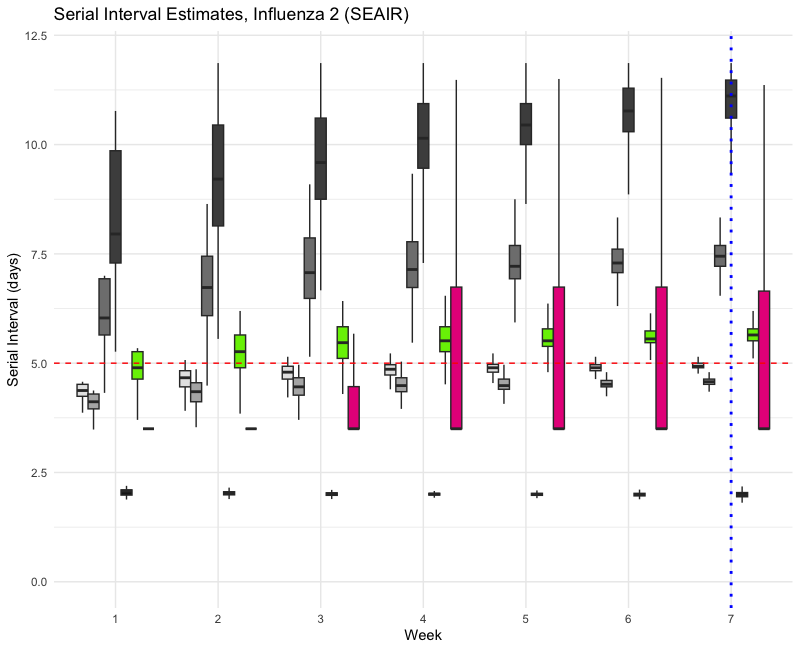}%
            
        } \\
        \vspace{1em}
        \includegraphics[width=0.2\linewidth]{legend.pdf}
        \caption{Comparison of estimates of the serial interval (SI) from the SIR, SEIR and SEAIR datasets of both Influenza 1 and 2. Methods include the White and Pagano method, along with the proposed sequential Bayes method under the varying mispecifications of the prior listed in Table~\ref{tab: misspecifiedCases}.} 
        \label{SI_Influenza}
\end{figure}

Fig~\ref{SI_Influenza} similarly displays the estimation results of the $SI$ for the same six datasets. The well-specified models are shown in Fig~\ref{SI_Influenza}(a) and (b), and immediately we see that the estimation of the SI is much more sensitive to misspecification compared to the estimation of $R_0$. The White and Pagano estimates do not change throughout the weeks, but the variance of these estimates does increase in Week 4, where the boxplot intercepts the true value of the SI in Weeks 4, 5, and 6. The well-specified sequential Bayes estimates, as well as the sequential Bayes misspecifications 1 and 2, perform very well, the latter two cases approaching the true value of the SI as more data is accumulated. The variance of these three cases, as well as the bias, are very small. Misspecification 3, in both Influenza 1 and 2, correctly estimates the SI in Week 1, but then overestimates and approaches an estimate of 7.5 days, which is 2.5 days greater than the true value of the SI. We note that, in this specific prior misspecification, we overestimated the SI by 1.5 days, as well as overestimated $R_0$ by 0.5. In Misspecifications 4 and 5, which were the two worst performing scenarios in the sensitivity analysis, we fixed the $R_0$ to be 1.33 greater than the true value, then underestimated the SI by 3 days and overestimated the SI by 3 days, respectively. The estimates found under Misspecification 4 have a small variance, but do not change from approximately 2 days as more data is added. Conversely, the estimates found under Misspecification 5 do note approach the true value of the SI, but rather continue to increase as more data is added, resulting in a much larger bias compared to the other estimates. 

Figure \ref{SI_Influenza}(c)-(f) show the results under a misspecified model, specifically under SEIR and SEAIR models. The results are very similar to what was seen in the case of the well-specfied models: the SI is much more sensitive to misspecifications in the prior compared to the $R_0$. The well-specified sequential Bayes estimates, as well as the sequential Bayes estimates found under Misspecifications 1 and 2, perform well, having the smallest variance and bias compared to the other estimates. The White and Pagano estimates provide more information in the Influenza 1-SEIR and SEAIR datasets compared to what was observed in the well-specified case, however this method results in performance that is comparable to the estimates found under the sequential Bayes Misspecification 3, and does not perform better than the aforementioned sequential Bayes methods. Again, the worst performing cases in the sensitivity analysis are those that have the worst bias in this simulation study: Misspecifications 4 and 5. 

\begin{figure}
    \centering
        \subfloat[Influenza 1, SIR]{%
            \includegraphics[width=.4\linewidth]{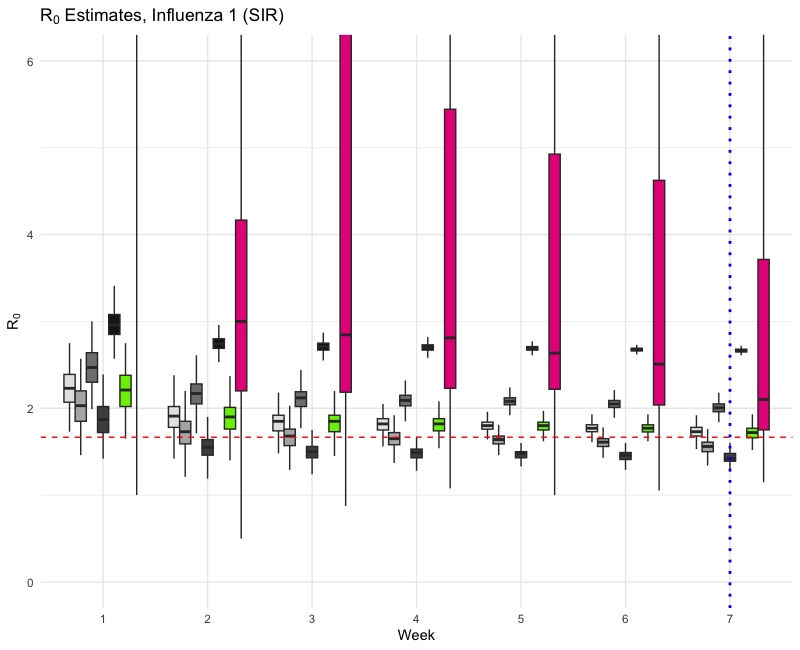}%
            
        } 
        \subfloat[Influenza 2, SIR]{%
            \includegraphics[width=.4\linewidth]{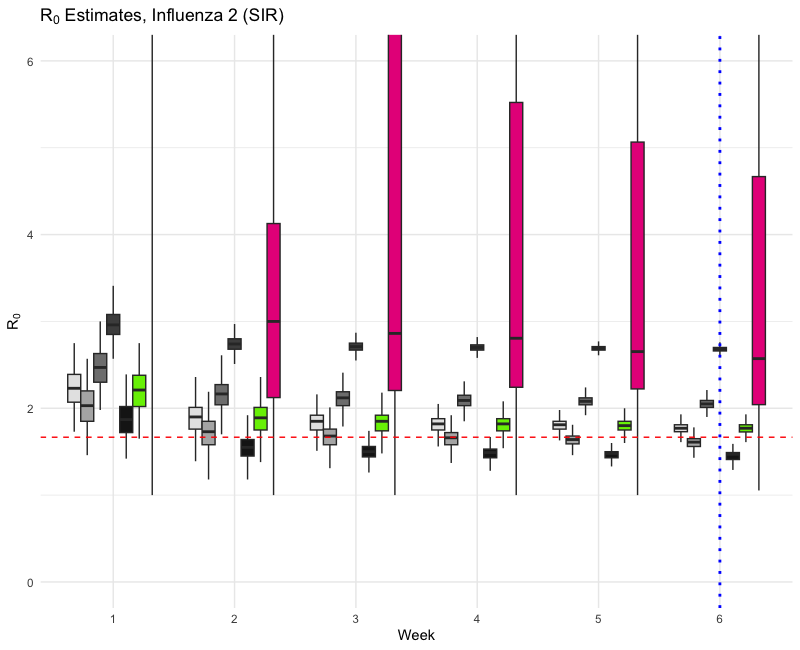}%
            
        } \\ 
        \vspace{1em} 
        \subfloat[Influenza 1, SEIR]{%
            \includegraphics[width=.4\linewidth]{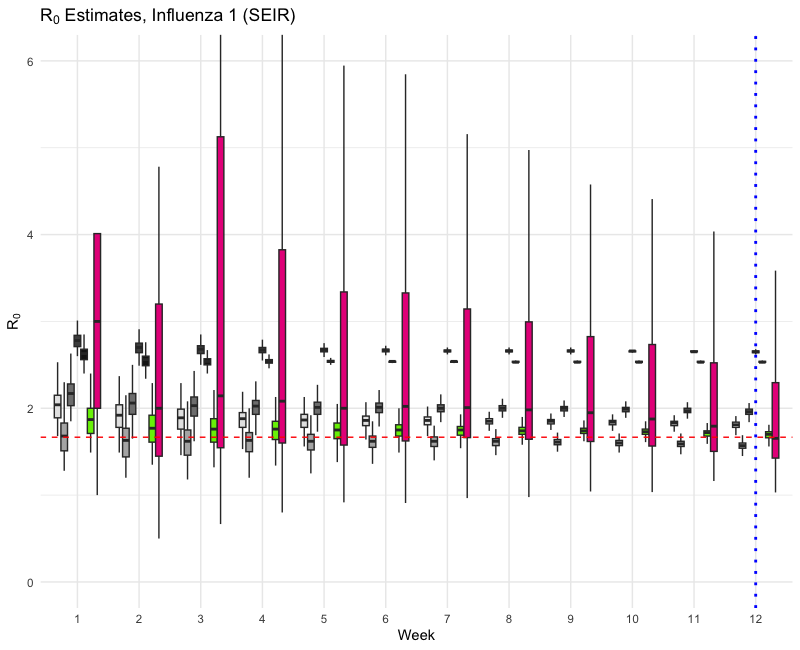}%
            
        }
        \subfloat[Influenza 2, SEIR]{%
            \includegraphics[width=.4\linewidth]{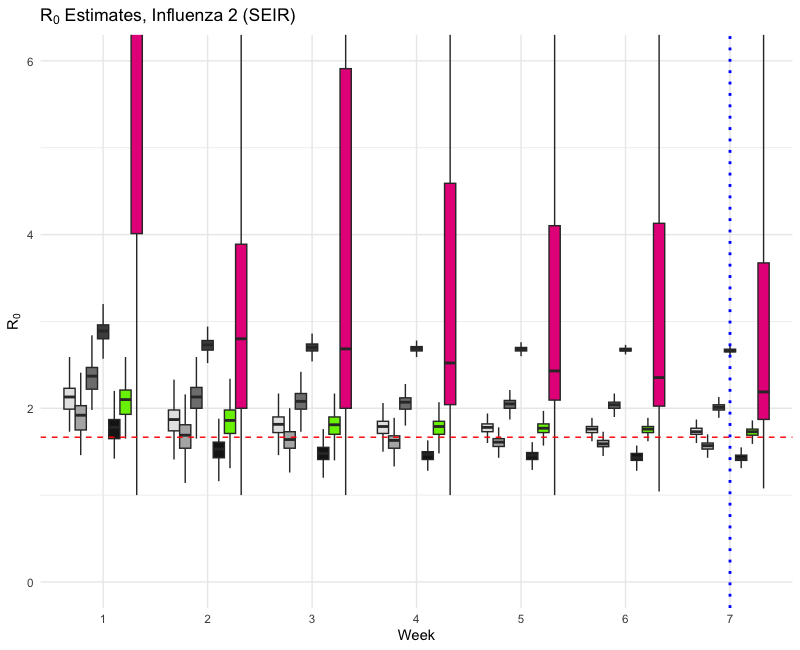}%
            
        } \\ 
        \vspace{1em} 
        \subfloat[Influenza 1, SEAIR]{%
            \includegraphics[width=.4\linewidth]{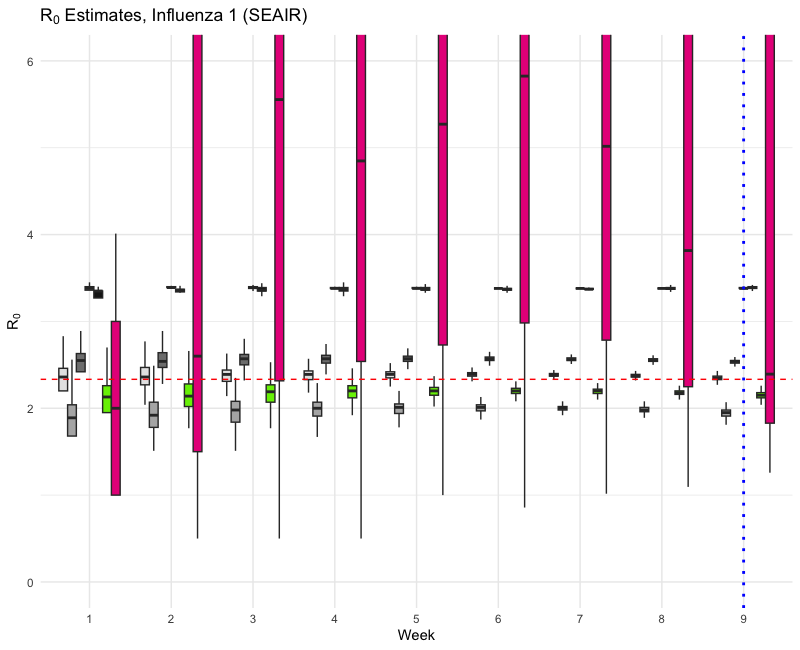}%
            
        } 
        \subfloat[Influenza 2, SEAIR]{%
            \includegraphics[width=.4\linewidth]{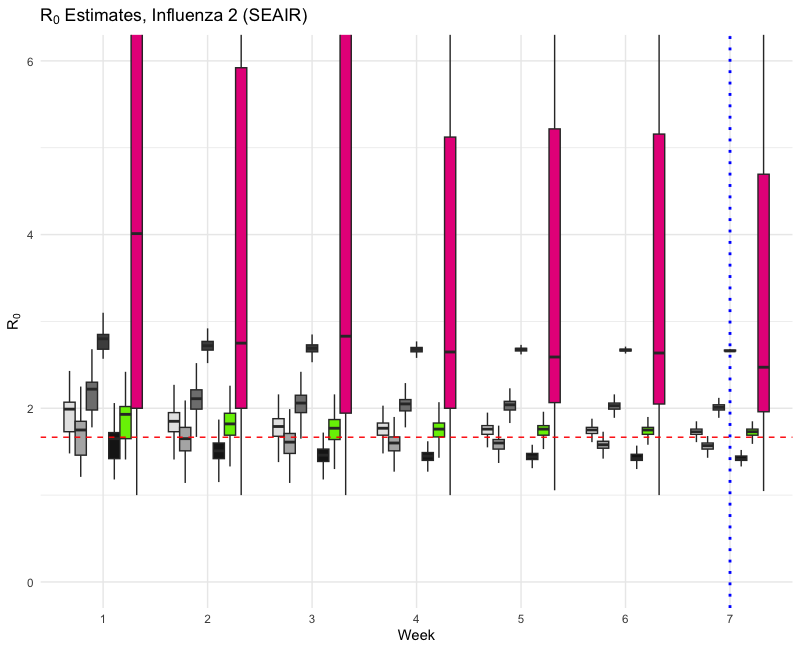}%
            
        } \\
        \vspace{1em}
        \includegraphics[width=0.2\linewidth]{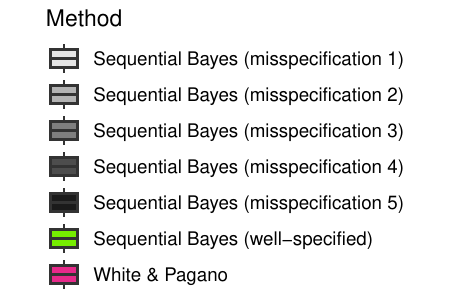}
        \caption{Comparison of estimates of $R_0$ from the SIR, SEIR, and SEAIR datasets of both Influenza 1 and 2. Methods include the White and Pagano method, along with the proposed sequential Bayes method under the varying mispecifications of the prior listed in Table~\ref{tab: misspecifiedCases}.} 
        \label{R0_Influenza}
    \end{figure}

\section{Example} 
\label{s: realExample}

To demonstrate our proposed sequential Bayes joint estimator on real data, we use Canadian COVID-19 data collected daily from January 25th, 2020 to December 7th, 2021. This was the same dataset used by Boonpatcharanon et al. \cite{boonpatcharanon2022estimating}, where they found the sequential Bayes estimator to be one of the better performing estimators in the six that they compared. We now compare how our joint estimator compares to the White and Pagano method. 

\begin{figure}
    \centering
        \subfloat[Canada]{%
            \includegraphics[width=.5\linewidth]{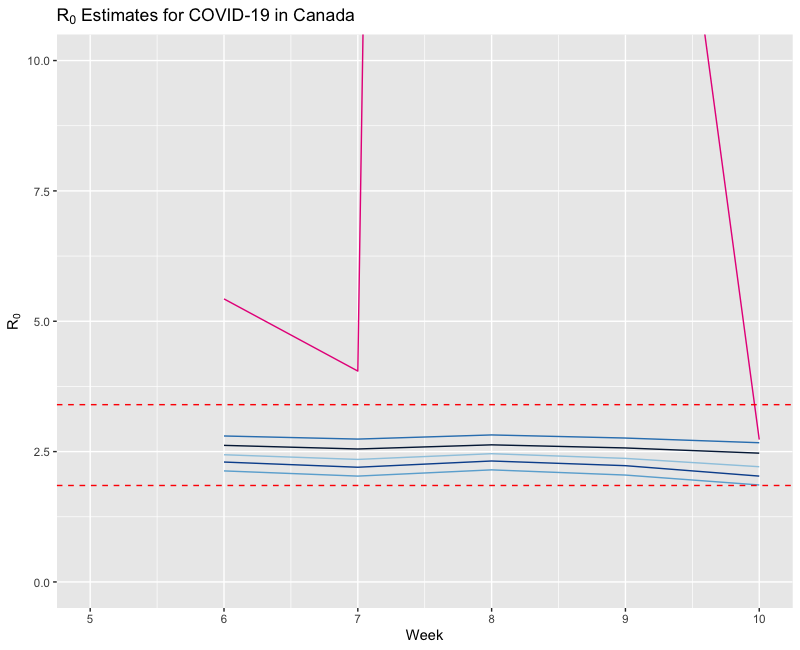}%
            
        } 
        \subfloat[British Columbia]{%
            \includegraphics[width=.5\linewidth]{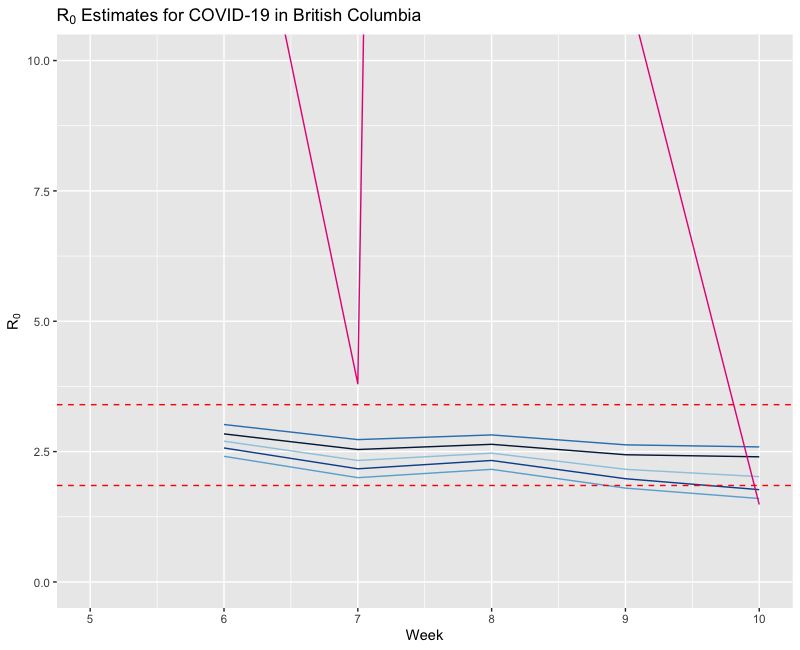}%
            
        } \\
        \vspace{1em} 
        \subfloat[Ontario]{%
            \includegraphics[width=.5\linewidth]{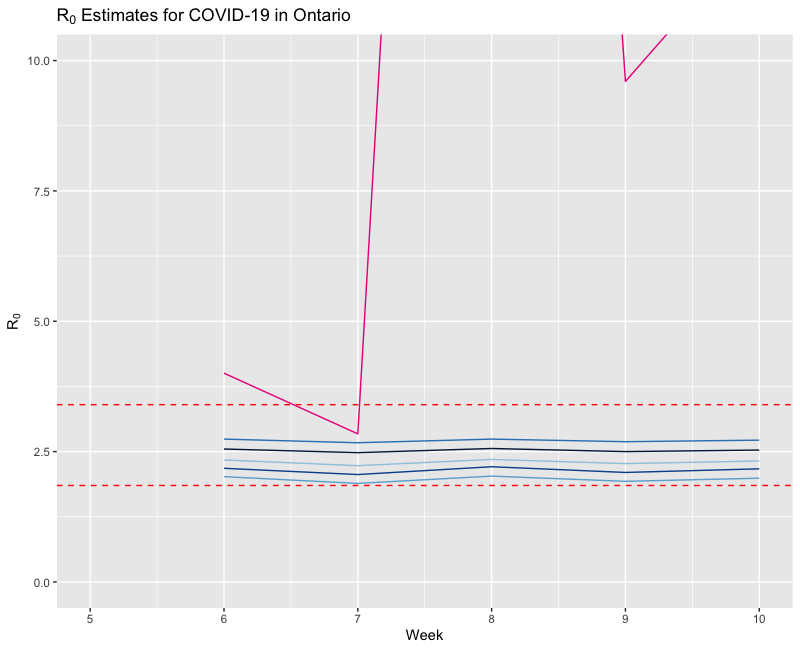}%
            
        }
        \subfloat[Quebec]{%
            \includegraphics[width=.5\linewidth]{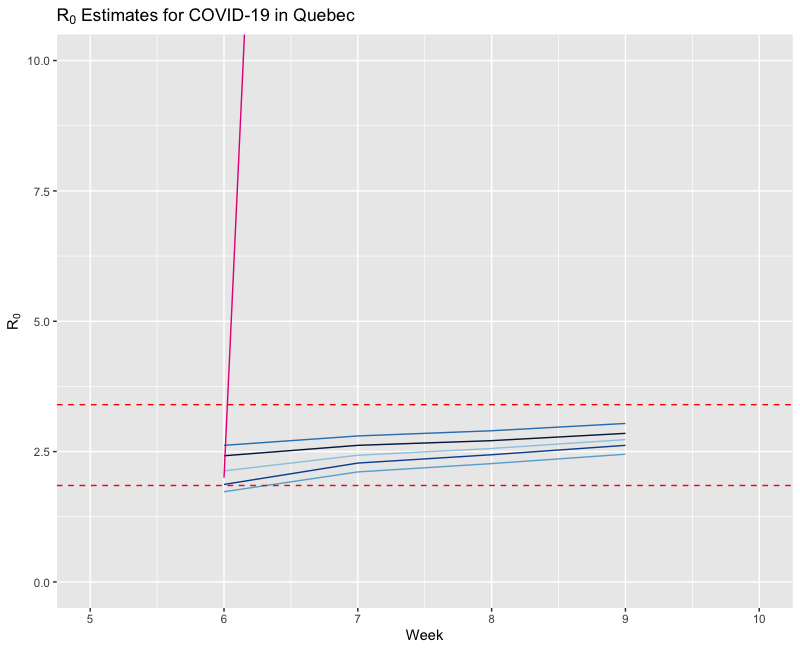}%
            
        } \\ 
        \vspace{1em} 
        \includegraphics[width=0.3\linewidth]{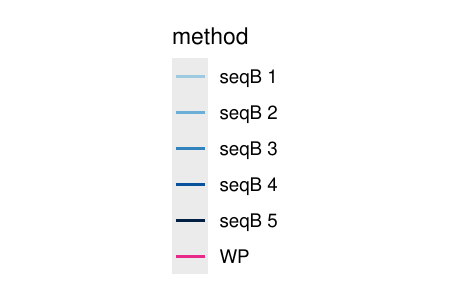}
        \caption{Comparison of $R_0$ estimates for Canadian COVID-19 data found using the White and Pagano method, along with our proposed sequential Bayes method using prior specifications outlined in Table~\ref{tab: examplePriors}} 
        \label{Example-R0}
\end{figure}

\begin{figure}
    \centering
        \subfloat[Canada]{%
            \includegraphics[width=.5\linewidth]{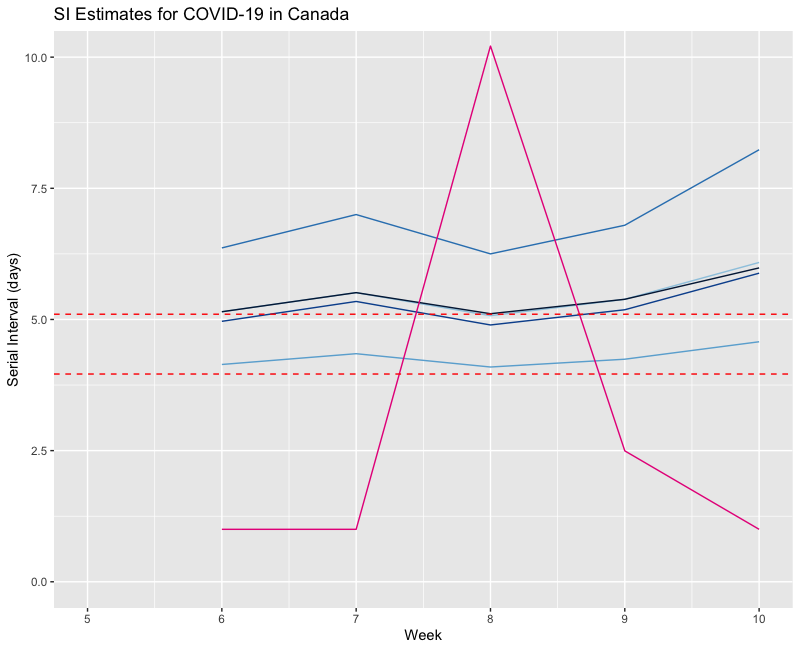}%
            
        } 
        \subfloat[British Columbia]{%
            \includegraphics[width=.5\linewidth]{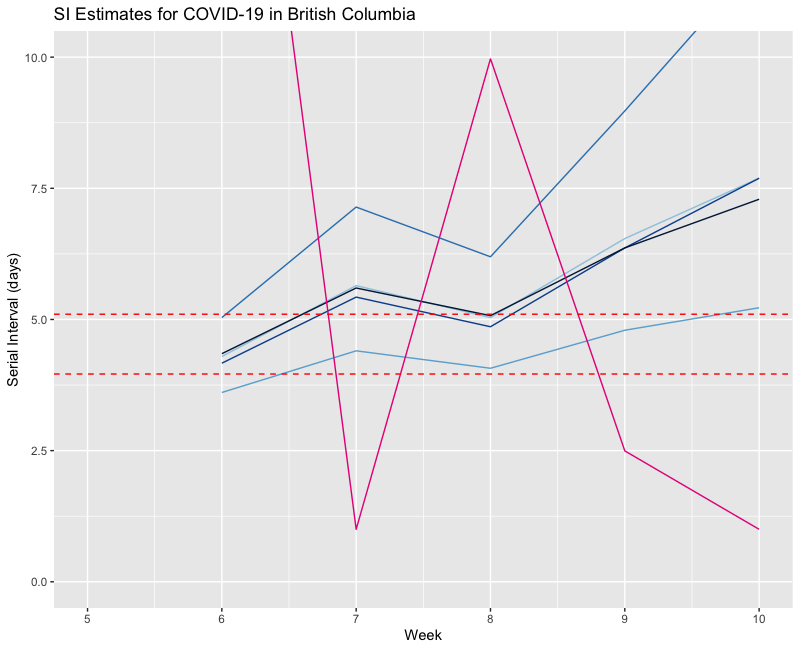}%
            
        } \\
        \vspace{1em} 
        \subfloat[Ontario]{%
            \includegraphics[width=.5\linewidth]{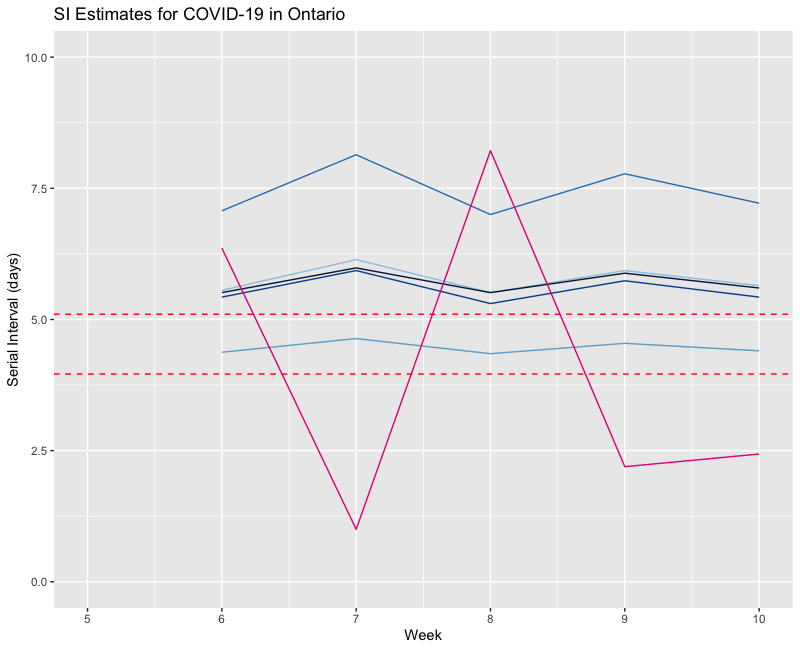}%
            
        }
        \subfloat[Quebec]{%
            \includegraphics[width=.5\linewidth]{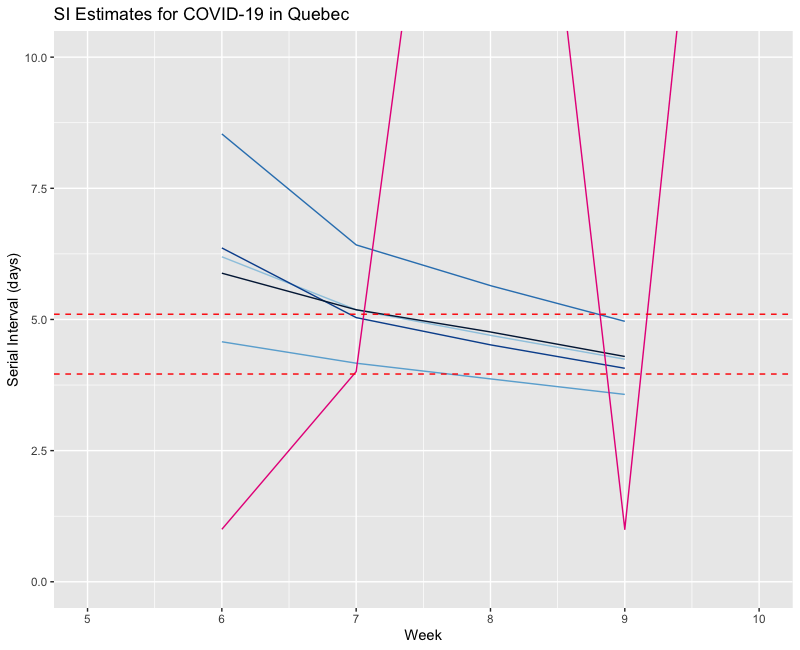}%
            
        } \\ 
        \vspace{1em} 
        \includegraphics[width=0.3\linewidth]{legend_example.pdf}
        \caption{Comparison of SI estimates for Canadian COVID-19 data found using the White and Pagano method, along with our proposed sequential Bayes method using prior specifications outlined in Table~\ref{tab: examplePriors}} 
        \label{Example-SI}
\end{figure}

We test our estimator for all of Canada, and for British Columbia, Ontario, and Quebec - the three most populous provinces. We compare the resulting estimates to previously found estimates of $R_0$ and the SI (red, dotted horizontal lines) \cite{knight2020estimating}.  We test the sequential Bayes estimator under five different priors listed in Table~\ref{tab: examplePriors}. The resulting estimates of $R_0$ and the SI are shown in Figures~\ref{Example-R0} and \ref{Example-SI}, respectively. We note that although we are representing the results as a line graph, the estimates are discrete and correspond to the specific weeks considered, here Weeks 6 to 10. We connect the points with lines solely to aid visual interpretation. From Figure~\ref{Example-R0}, we see that not only are sequential Bayes estimates of $R_0$ much more stable than the White and Pagano estimates, but also that the sequential Bayes estimates are not very sensitive to specifications in the prior. Under the varying prior configurations we tested, the resulting $R_0$ estimates do not change much, and remain similar to the benchmark estimates we use as a comparison. As expected, from what was concluded in the simulation study, the estimates of the SI are much more sensitive to changes in the prior specification, as shown in Figure~\ref{Example-SI}. We do note, however, that although the sequential Bayes estimates are sensitive, they are more stable than the White and Pagano estimates, which tend to vary, sometimes greatly as is the case for the Quebec dataset, week to week.  

\begin{table}[h!]
\centering
\begin{tabular}{lcc}
\toprule
\textbf{Case} & $SI$ & $R_0$ \\
\midrule
seqB 1 & $5$ & $2.5$ \\
seqB 2 & $4$    & $2$      \\
seqB 3 & $6$   & $3$     \\
seqB 4 & $6$   & $2$ \\
seqB 5 & $4$   & $3$     \\
\bottomrule
\end{tabular}
\caption{Prior specifications used in the real data analysis on the Canadian COVID-19 dataset. }
\label{tab: examplePriors}
\end{table}

\section{Discussion}

Estimating both the basic reproduction number, $R_0$, and the serial interval (SI) at the beginning of an infectious disease outbreak provides valuable information that can inform public health measures aimed at reducing transmission. Although many estimators exist for these two parameters when considered separately, to our knowledge there is only one method that jointly estimates them using case count data \cite{forsberg2008likelihood}. In this work, we have proposed a joint estimator using the sequential Bayes method to estimate $R_0$ and the SI simultaneously from case counts. Our estimator assumes the data arise from an SIR model; however, through a comprehensive set of simulation studies, we have shown that even under model misspecification, and under substantial prior misspecification, our estimator performs well compared to the one other known joint estimation method.

Our joint sequential Bayes method uses a copula to define the joint prior, where the marginal distributions of $R_0$ and the SI are specified as log-Gamma and joined through a Gaussian copula. Using this prior, we then could implement our sequential Bayes method, where the prior updates at each time point to be the posterior of the previous time point. In our case, since we were working with weekly data, the prior at Week $i+1$ was the posterior found at Week $i$, $i = 1, ..., n$. We confirmed through simulation that as data is added, the posterior behaves as expected and begins to resemble the theoretical curve describing the relationship between $R_0$ and the SI using infectious disease counts. This confirms that our prior specification does not dominate and that the data appropriately drive the estimates as additional weeks are observed.

We also conducted a sensitivity analysis to assess how sensitive the two parameters are to misspecification in the prior, specifically, to misspecification of the marginal means of $R_0$ and the SI. The results, shown in Figure~\ref{fig: intensiveSimulation}, show that the SI is much more sensitive to misspecification compare to the $R_0$, with the latter remaining robust. We used this sensitivity analysis to select five representative scenarios to explore further in the following simulation study, where we evaluated our method not only under prior misspecification but also under model misspecification, using data simulated from SIR, SEIR, and SEAIR models. Representing the results as boxplots, Figures~\ref{R0_Influenza} and \ref{SI_Influenza} display the estimates of $R_0$ and SI, respectively, comparing those found through our sequential Bayes method and the White and Pagano method.  

Overall, and as expected from the sensitivity analysis, the sequential Bayes $R_0$ estimates were not very sensitive to misspecifications in the prior. Only in worst case scenarios, where we over-specified both the SI and $R_0$, did we see large deviations from the true value. Even in these cases, at least in the early weeks of an infectious disease spread, these estimates still performed better in terms of bias and variance than those using the White and Pagano method.  Although the estimates of the SI were much more sensitive compared to the $R_0$ estimates, this simulation study gave us information on how the SI behaves. The White and Pagano method does not give much information about the serial interval, as in many cases considered the differences in estimates from week to week were very small. In contrast, under misspecification, the estimates using the sequential Bayes method can be misleading, with some cases highly overestimating the parameter. However, when misspecification is modest (within $\pm 1$ day), our sequential Bayes method results in quality estimates of the serial interval. Thus, if some prior information is known of the serial interval, either from other regions or similar infectious diseases, then our proposed method performs well in estimating the SI. If there is no prior information known, then we suggest using both our method, along with the White and Pagano method, to compare the resulting estimates of SI to make a decision. When estimating $R_0$, our method outperformed, or had very comaprable results, to the White and Pagano method in the early weeks of a simulated infectious disease in all cases. 

We tested our method on real Canadian COVID-19 data as well, and found similar results to what was seen in the simulation studies. When estimating $R_0$, the results shown in Figure~\ref{Example-R0}, estimates using our proposed sequential Bayes estimator under multiple prior specifications was much more stable compared to estimates found using the White and Pagano method which varied substantially week to week. Further, the estimates were not sensitive to changes in the prior specification. Although estimates of the SI were sensitive to changes in the prior specification, as shown in Figure~\ref{Example-SI}, again, the estimates found were more stable than the White and Pagano estimates. We encourage researchers estimating the $R_0$ and the SI, to use multiple estimators and to assess robustness under misspecification, as we did in this real data analysis.

In this work, we provide an estimator which is able to estimate both the $R_0$ and the SI jointly using mildly informative priors in a sequential and interpretable framework. As future work, we plan to investigate whether having daily data leads to less sensitive SI estimates. Although weekly case counts are more commonly available during outbreaks, daily counts may be accessible early on when cases are being closely monitored. If daily data lead to greater stability, our method could provide high-quality estimates even earlier in an outbreak. Additionally, we aim to extend this work to allow for multi-scale estimation, particularly the use of data from a larger region to obtain reliable estimates of both the $R_0$ and the SI for smaller subregions. 

\clearpage

\bibliographystyle{apalike}

\begin{thebibliography}{10}

\bibitem{boonpatcharanon2022estimating}
Boonpatcharanon S, Heffernan JM, Jankowski H.
\newblock Estimating the basic reproduction number at the beginning of an outbreak.
\newblock Plos one. 2022;17(6):e0269306.

\bibitem{svensson2007note}
Svensson {\AA}.
\newblock A note on generation times in epidemic models.
\newblock Mathematical biosciences. 2007;208(1):300--311.

\bibitem{britton2019estimation}
Britton T, Scalia~Tomba G.
\newblock Estimation in emerging epidemics: biases and remedies.
\newblock Journal of the Royal Society Interface. 2019;16(150):20180670.

\bibitem{park2021forward}
Park SW, Sun K, Champredon D, Li M, Bolker BM, Earn DJ, et~al.
\newblock Forward-looking serial intervals correctly link epidemic growth to reproduction numbers.
\newblock Proceedings of the National Academy of Sciences. 2021;118(2):e2011548118.

\bibitem{cori2013new}
Cori A, Ferguson NM, Fraser C, Cauchemez S.
\newblock A new framework and software to estimate time-varying reproduction numbers during epidemics.
\newblock American journal of epidemiology. 2013;178(9):1505--1512.

\bibitem{wallinga2004different}
Wallinga J, Teunis P.
\newblock Different epidemic curves for severe acute respiratory syndrome reveal similar impacts of control measures.
\newblock American Journal of epidemiology. 2004;160(6):509--516.

\bibitem{zhao2020estimating}
Zhao S, Gao D, Zhuang Z, Chong MK, Cai Y, Ran J, et~al.
\newblock Estimating the serial interval of the novel coronavirus disease (COVID-19): a statistical analysis using the public data in Hong Kong from January 16 to February 15, 2020.
\newblock Frontiers in Physics. 2020;8:347.

\bibitem{talmoudi2020estimating}
Talmoudi K, Safer M, Letaief H, Hchaichi A, Harizi C, Dhaouadi S, et~al.
\newblock Estimating transmission dynamics and serial interval of the first wave of COVID-19 infections under different control measures: a statistical analysis in Tunisia from February 29 to May 5, 2020.
\newblock BMC infectious diseases. 2020;20:1--8.

\bibitem{griffin2020rapid}
Griffin J, Casey M, Collins {\'A}, Hunt K, McEvoy D, Byrne A, et~al.
\newblock Rapid review of available evidence on the serial interval and generation time of COVID-19.
\newblock BMJ open. 2020;10(11):e040263.

\bibitem{cowling2009estimation}
Cowling BJ, Fang VJ, Riley S, Peiris JSM, Leung GM.
\newblock Estimation of the serial interval of influenza.
\newblock Epidemiology. 2009;20(3):344--347.

\bibitem{forsberg2008likelihood}
Forsberg~White L, Pagano M.
\newblock A likelihood-based method for real-time estimation of the serial interval and reproductive number of an epidemic.
\newblock Statistics in medicine. 2008;27(16):2999--3016.

\bibitem{nishiura2020serial}
Nishiura H, Linton NM, Akhmetzhanov AR.
\newblock Serial interval of novel coronavirus (COVID-19) infections.
\newblock International journal of infectious diseases. 2020;93:284--286.

\bibitem{thompson2019improved}
Thompson RN, Stockwin JE, van Gaalen RD, Polonsky JA, Kamvar ZN, Demarsh PA, et~al.
\newblock Improved inference of time-varying reproduction numbers during infectious disease outbreaks.
\newblock Epidemics. 2019;29:100356.

\bibitem{ganyani2020estimating}
Ganyani T, Kremer C, Chen D, Torneri A, Faes C, Wallinga J, et~al.
\newblock Estimating the generation interval for coronavirus disease (COVID-19) based on symptom onset data, March 2020.
\newblock Eurosurveillance. 2020;25(17):2000257.

\bibitem{donnelly2011serial}
Donnelly CA, Finelli L, Cauchemez S, Olsen SJ, Doshi S, Jackson ML, et~al.
\newblock Serial intervals and the temporal distribution of secondary infections within households of 2009 pandemic influenza A (H1N1): implications for influenza control recommendations.
\newblock Clinical Infectious Diseases. 2011;52(suppl\_1):S123--S130.

\bibitem{becker2010type}
Becker N, Wang D, Clements M.
\newblock Type and quantity of data needed for an early estimate of transmissibility when an infectious disease emerges.
\newblock Eurosurveillance. 2010;15(26):19603.

\bibitem{moser2015impact}
Moser CB, Gupta M, Archer BN, White LF.
\newblock The impact of prior information on estimates of disease transmissibility using Bayesian tools.
\newblock PloS one. 2015;10(3):e0118762.

\bibitem{bettencourt2008real}
Bettencourt LM, Ribeiro RM.
\newblock Real time bayesian estimation of the epidemic potential of emerging infectious diseases.
\newblock PloS one. 2008;3(5):e2185.

\bibitem{yang2013bayesian}
Yang F, Yuan L, Tan X, Huang C, Feng J.
\newblock Bayesian estimation of the effective reproduction number for pandemic influenza A H1N1 in Guangdong Province, China.
\newblock Annals of epidemiology. 2013;23(6):301--306.

\bibitem{narula2015bayesian}
Narula P, Azad S, Lio P.
\newblock Bayesian melding approach to estimate the reproduction number for tuberculosis transmission in Indian states and union territories.
\newblock Asia Pacific Journal of Public Health. 2015;27(7):723--732.

\bibitem{guo2023estimating}
Guo Z, Zhao S, Yam CHK, Li C, Jiang X, Chow TY, et~al.
\newblock Estimating the serial intervals of SARS-CoV-2 Omicron BA. 4, BA. 5, and BA. 2.12. 1 variants in Hong Kong.
\newblock Influenza and Other Respiratory Viruses. 2023;17(2):e13105.

\bibitem{elderd2013population}
Elderd BD, Dwyer G, Dukic V.
\newblock Population-level differences in disease transmission: A Bayesian analysis of multiple smallpox epidemics.
\newblock Epidemics. 2013;5(3):146--156.

\bibitem{fisman2013idea}
Fisman DN, Hauck TS, Tuite AR, Greer AL.
\newblock An IDEA for short term outbreak projection: nearcasting using the basic reproduction number.
\newblock PloS one. 2013;8(12):e83622.

\bibitem{knight2020estimating}
Knight J, Mishra S.
\newblock Estimating effective reproduction number using generation time versus serial interval, with application to COVID-19 in the Greater Toronto Area, Canada.
\newblock Infectious Disease Modelling. 2020;5:889--896.

\end{thebibliography}

\clearpage

\section*{Supporting Information}

\subsection*{S1: Conjugate Prior} 

From Eq \ref{eq: infections}, we assume that the conditional distribution of $I(t_{j+1})$, conditional on both $I(t_j)$, $R_0$, and $\gamma$ is Poisson with mean $\lambda = I(t_j)\exp\left[(t_{j+1} - t_j)\gamma(R_0-1)\right]$. In other words, we assume that

$$I(t_{j+1})|I(t_j), R_0, \gamma \sim Poisson(\lambda).$$ 

Now, if we let $\theta = (t_{j+1} - t_j)\gamma(R_0-1)$, then we have

$$I(t_{j+1})|I(t_j), R_0, \gamma \sim Poisson(I(t_j)\exp[\theta]).$$

We now assume 

$$\theta \sim logGamma(\alpha, \beta),$$

and we show that the log-Gamma prior is the conjugate prior. 

We begin with the likelihood:

\begin{equation} \label{likelihood}
\begin{split}
L(I(t_{j+1})|\theta) & = \prod_{j=1}^n\left[\frac{\left(I(t_j)e^\theta\right)^{I(t_{j+1})}e^{-I(t_j)e^\theta}}{I(t_{j+1})!}\right] \\
 & \propto e^{\theta \sum_{j=1}^n I(t_{j+1})}e^{-e^\theta \sum_{j=1}^n I(t_j)}.
\end{split}
\end{equation}

Then, we use the following pdf of the logGamma distribution with shape parameter $\alpha$ and scale parameter $\beta$: 

$$\pi (\theta) = \frac{1}{\Gamma(\alpha)}\left[\frac{e^\theta}{\beta} \right]^\alpha e^{\frac{-e^\theta}{\beta}}.$$

Now, we derive the posterior. 

\begin{equation} \label{likelihood}
\begin{split}
\pi(\theta|I(t_{j+1})) & \propto \pi(\theta)L(I(t_{j+1})) \\ 
& \propto \frac{1}{\Gamma(\alpha)}\left[\frac{e^\theta}{\beta} \right]^\alpha e^{\frac{-e^\theta}{\beta}}e^{\theta \sum_{j=1}^n I(t_{j+1})}e^{-e^\theta \sum_{j=1}^n I(t_j)} \\
 & \propto e^{\alpha\theta}e^{\frac{-e^\theta}{\beta}}e^{\theta \sum_{j=1}^n I(t_{j+1})}e^{-e^\theta \sum_{j=1}^n I(t_j)} \\ 
 & = e^{(\alpha + \sum_{j=1}^n I(t_{j+1}))\theta}e^{-(\frac{1}{\beta} + \sum_{j=1}^n I(t_j))e^\theta}.
\end{split}
\end{equation} 

We have just shown that 
$$\theta|I(t_{j+1}) \sim logGamma(\alpha^*, \beta^*),$$ 
where $\alpha^* = \alpha + S$, $\beta^* = \frac{1}{T+1/\beta}$, and $S = \sum_{j=1}^n I(t_{j+1})$, $T = \sum_{j=1}^n I(t_{j})$. 

\subsection*{S2: Other Compartmental Models}

In the main text, we introduced the SIR model in detail as our proposed sequential Bayes estimator assumes the data comes from this compartmental model. However, this is one of the simplest compartmental models and in practice this assumption could be violated. Thus, we test our method to determine robustness against violations of this assumption by considering data that comes from either an SEIR or SEAIR model. We describe both in detail below. 

We begin with the \textit{SEIR model}, which accounts for a potential latency period where individuals have been infected but are not yet infectious. In this model, an additional compartment is added, the exposed compartment, and the SEIR model is a system of four ordinary differential equations: 

\begin{align*} 
\frac{dS}{dt} &=  -\beta \frac{S(t)}{N}I(t) \\ 
\frac{dE}{dt} &=  -\beta \frac{S(t)}{N}I(t) - \sigma E(t)\\
\frac{dI}{dt} &=  \sigma E(t)-\gamma I(t) \\ 
\frac{dR}{dt} &= \gamma I(t)
\end{align*}

In this extension of the SIR model, $E(t)$ consists of all individuals exposed to the disease but not yet infectious, and $\sigma$ is the progression rate from the exposed to infectious compartment. Here, $N = S(t) + E(t) + I(t) + R(t)$, $R_0 = \beta/\gamma$ and the $SI = 1/\gamma + 1/\sigma$. 

The second compartment model we will discuss in the \textit{SEAIR model}, which includes another compartment accounting for individuals who are asymptomatic yet still infectious. This model is a system of five ordinary differential equations: 

\begin{align*} 
\frac{dS}{dt} &=  -\beta \frac{S(t)}{N}I(t) \\ 
\frac{dE}{dt} &=  \beta \frac{S(t)}{N}I(t) - \sigma E(t)\\
\frac{dE}{dt} &=  \sigma E(t) - \rho A(t)\\
\frac{dI}{dt} &=  \rho A(t)-\gamma I(t) \\ 
\frac{dR}{dt} &= \gamma I(t)
\end{align*}

$A(t)$ consists of all asymptomatic individuals who are infectious, and $\rho$ is the rate at which those asymptomatic individuals become symptomatic. Here, $N = S(t) + E(t) + A(t) + I(t) + R(t)$, $R_0 = \beta/\gamma + \beta/\rho$ and the $SI = 1/\gamma + 1/\sigma$. 

\subsection*{S3: Illustrative Example of Posterior Behaviour} 

\begin{figure}
    \centering
    \includegraphics[width=1\linewidth]{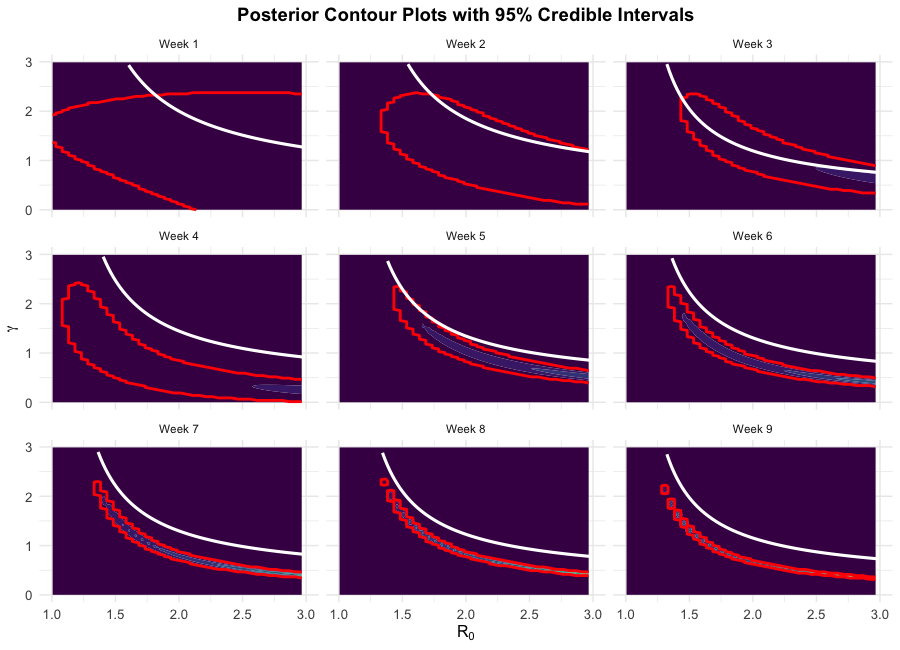}
    \caption{Each panel shows the prior that is used for that specific week represented by a contour plot, which, except for Week 1, is the posterior from the previous week. The density is shown such that the more vibrant colours indicate higher density regions. The red lines denote 95\% credible intervals (CrI) for the prior used in each week, which were found using the highest density regions (HDR). Specifically, within each CrI is 95\% of the density, i.e. $P((\gamma, R_0) \in CrI) = 0.95$.}
    \label{Supp-Posterior}
\end{figure}

In the main text, we included an example of how the posterior updates using our sequential Bayes estimator when the prior is well-specified, i.e. when the marginal means are equal to the true values of the parameters of interest. We tested this on the Influenza 1-SIR dataset. In Fig~\ref{Supp-Posterior}, we show how the posterior updates when the prior is misspecified using the same dataset. Specifically, we set the marginal means of $R_0$ and $SI$ to be 3 and 8 (days), respectively. As was seen in the well-specified case, even when the prior is misspecified, the data takes over at approximately Week 3 and the posterior begins to resemble the theoretical curve which models the relationship between $R_0$ and the SI. 

\subsection*{S4: Full Estimation Performance Plots} 

In the main text, we displayed the results from the estimation performance on a boxplot with reduced ranges for ease of visual interpretation. Here, we display the full range of boxplots for both the $R_0$ and the SI estimation under all six datasets considered. 

\begin{figure}
    \centering
        \subfloat[Influenza 1, SIR]{%
            \includegraphics[width=.4\linewidth]{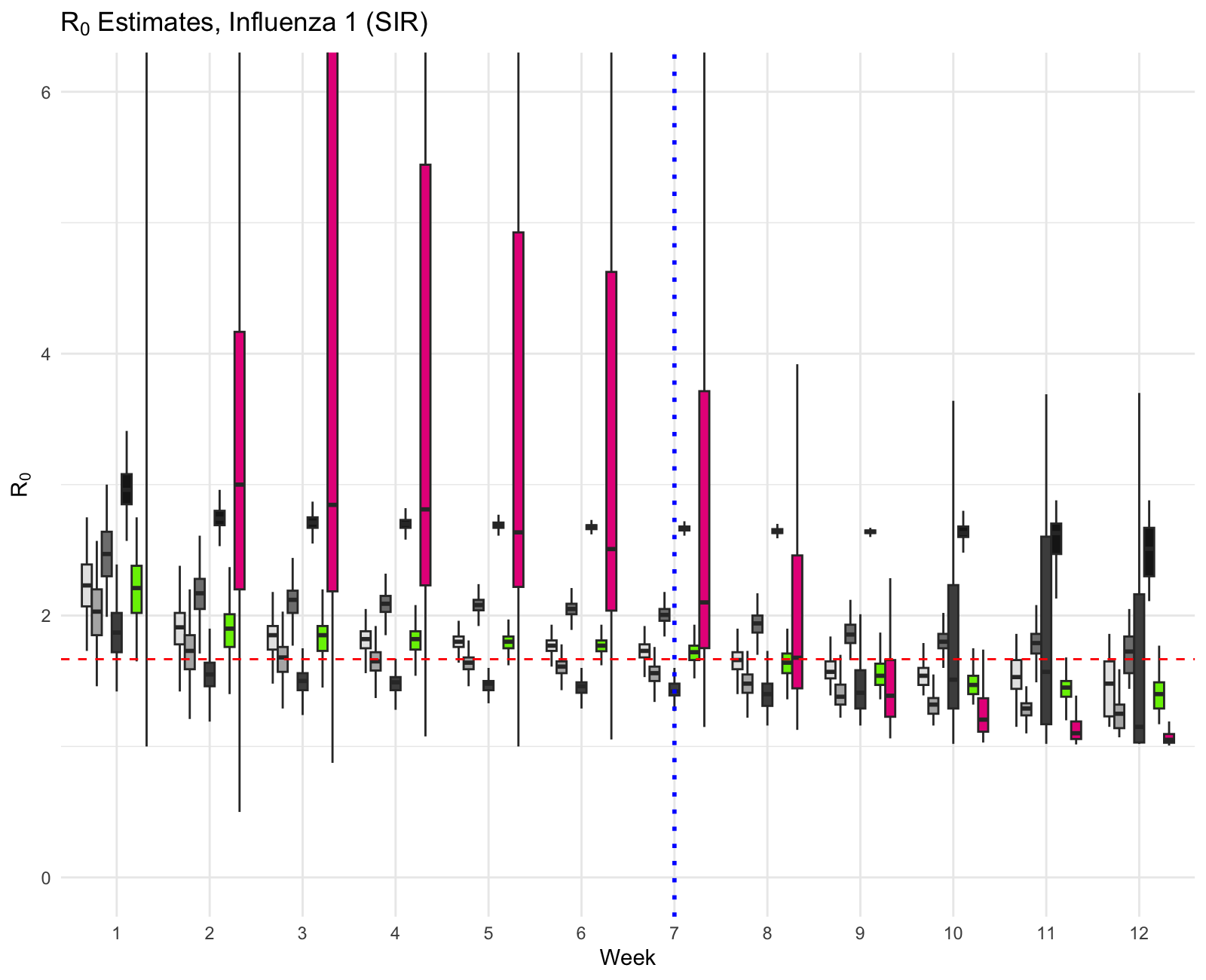}%
            
        } 
        \subfloat[Influenza 2, SIR]{%
            \includegraphics[width=.4\linewidth]{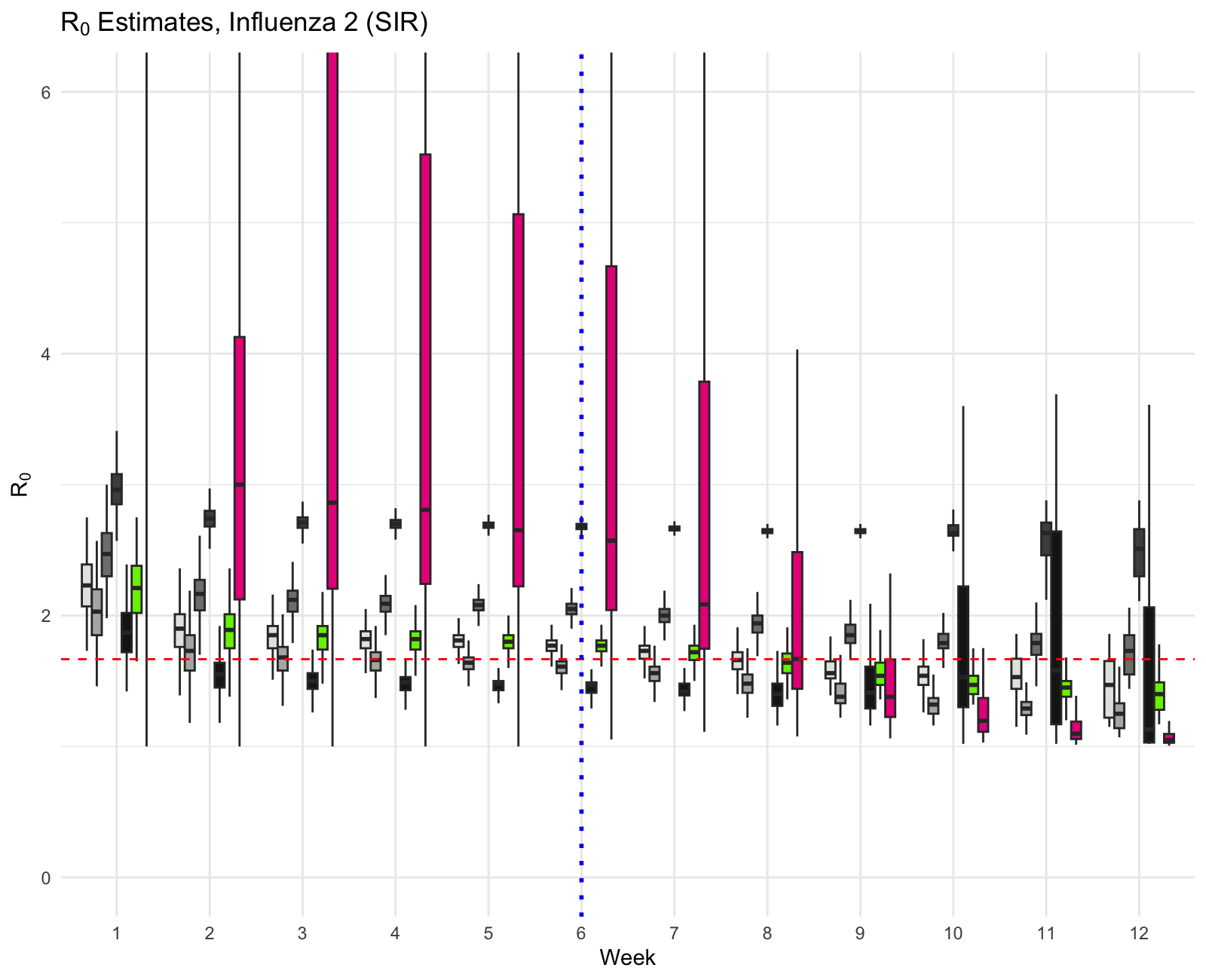}%
            
        } \\ 
        \vspace{1em} 
        \subfloat[Influenza 1, SEIR]{%
            \includegraphics[width=.4\linewidth]{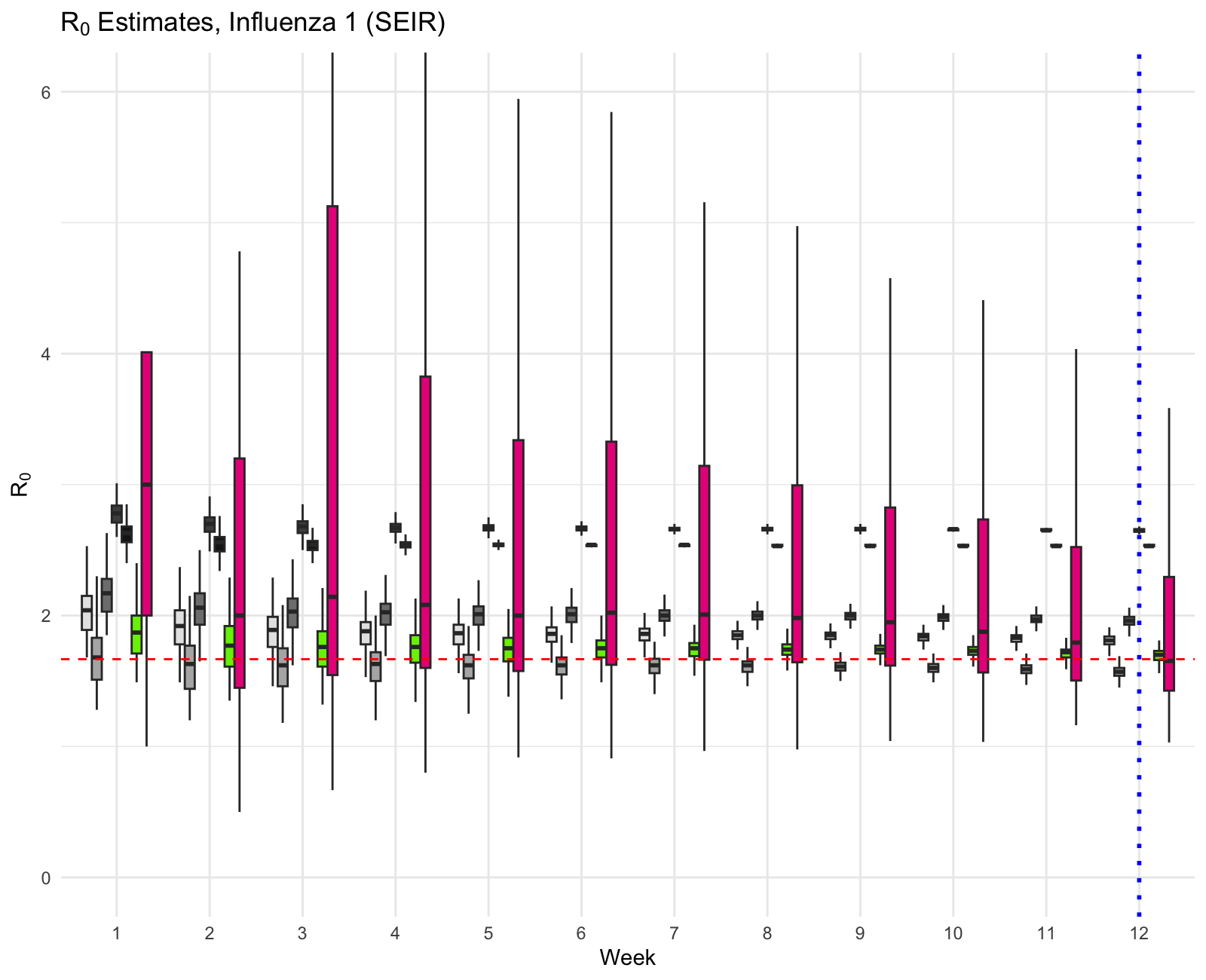}%
            
        }
        \subfloat[Influenza 2, SEIR]{%
            \includegraphics[width=.4\linewidth]{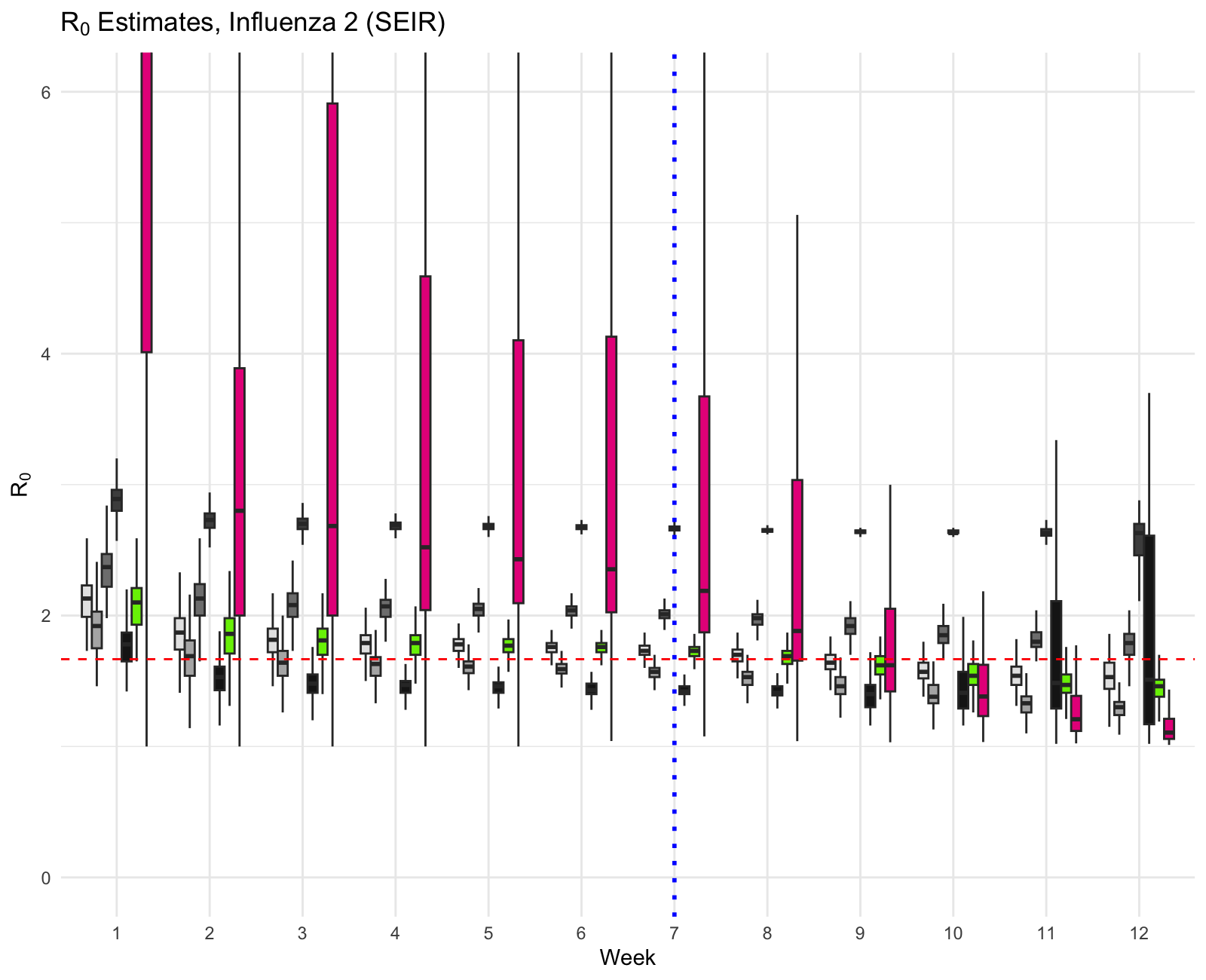}%
            
        } \\ 
        \vspace{1em} 
        \subfloat[Influenza 1, SEAIR]{%
            \includegraphics[width=.4\linewidth]{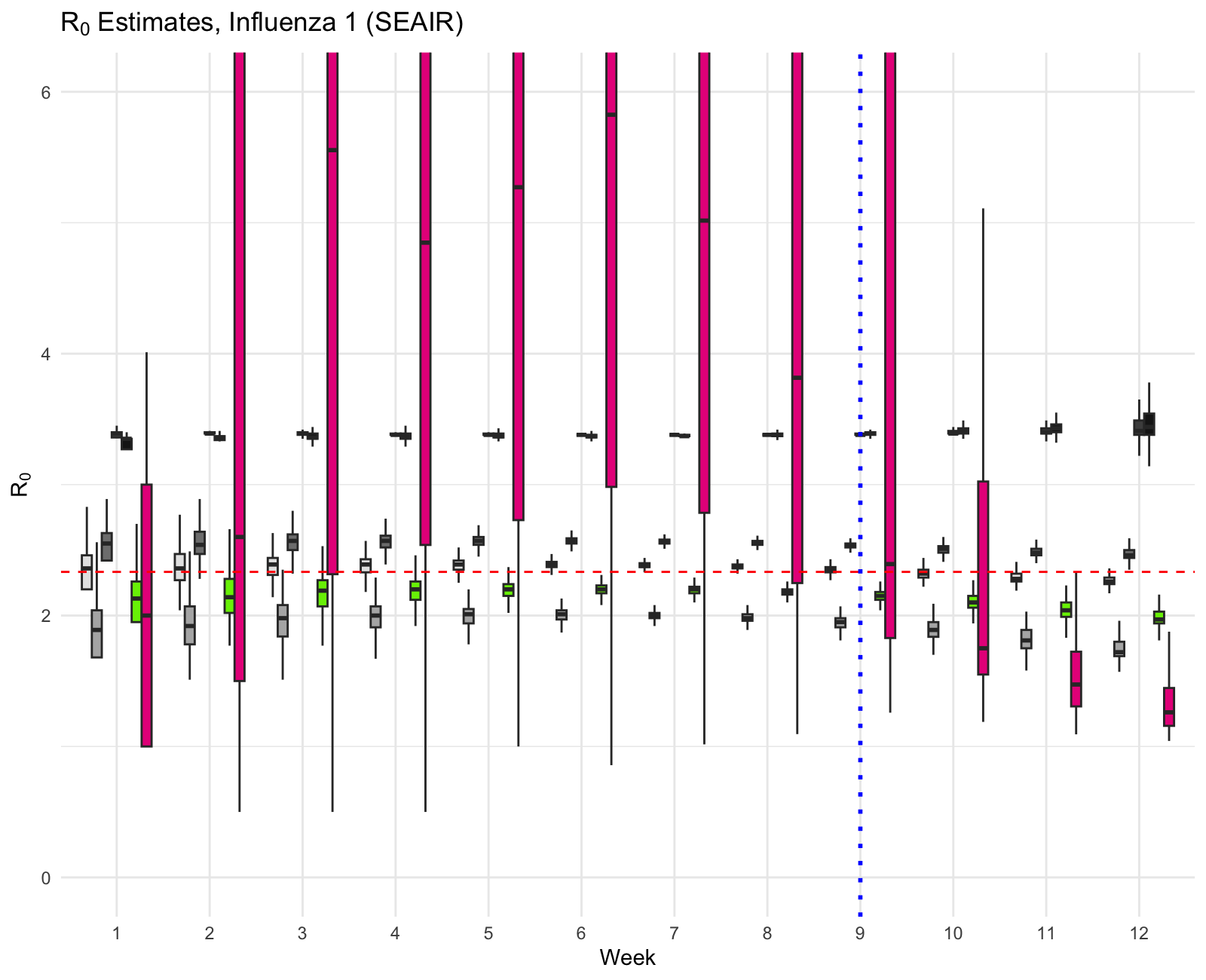}%
            
        } 
        \subfloat[Influenza 2, SEAIR]{%
            \includegraphics[width=.4\linewidth]{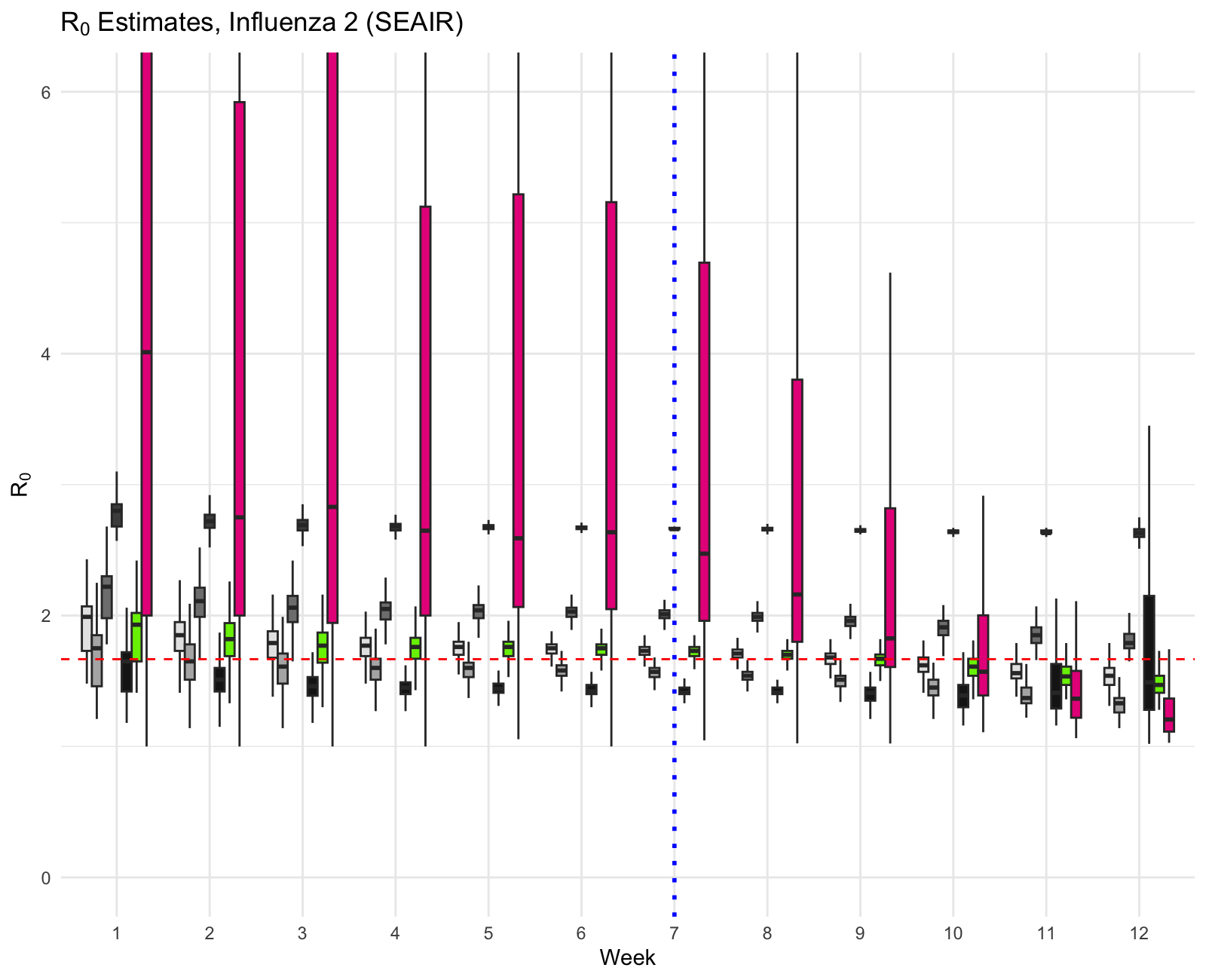}%
            
        } \\
        \vspace{1em}
        \includegraphics[width=0.2\linewidth]{legend.pdf}
        \caption{Comparison of estimates of $R_0$ from the SIR, SEIR, and SEAIR datasets of both Influenza 1 and 2. Methods include the White and Pagano method, along with the proposed sequential Bayes method under the varying mispecifications of the prior listed in Table~\ref{tab: misspecifiedCases}.} 
        \label{R0_Influenza-Supp}
    \end{figure}

    \begin{figure}
    \centering
        \subfloat[Influenza 1, SIR]{%
            \includegraphics[width=.4\linewidth]{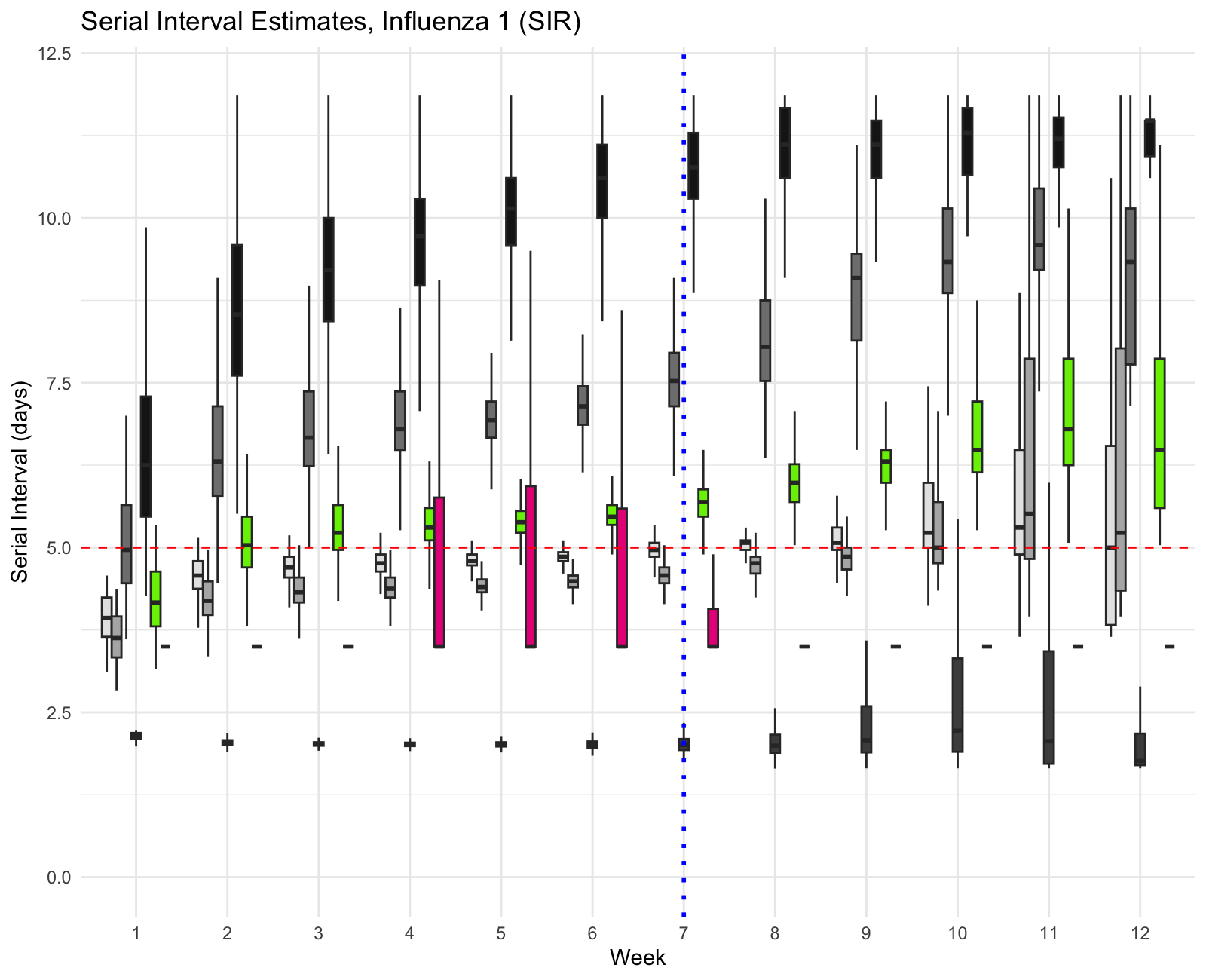}%
            
        } 
        \subfloat[Influenza 2, SIR]{%
            \includegraphics[width=.4\linewidth]{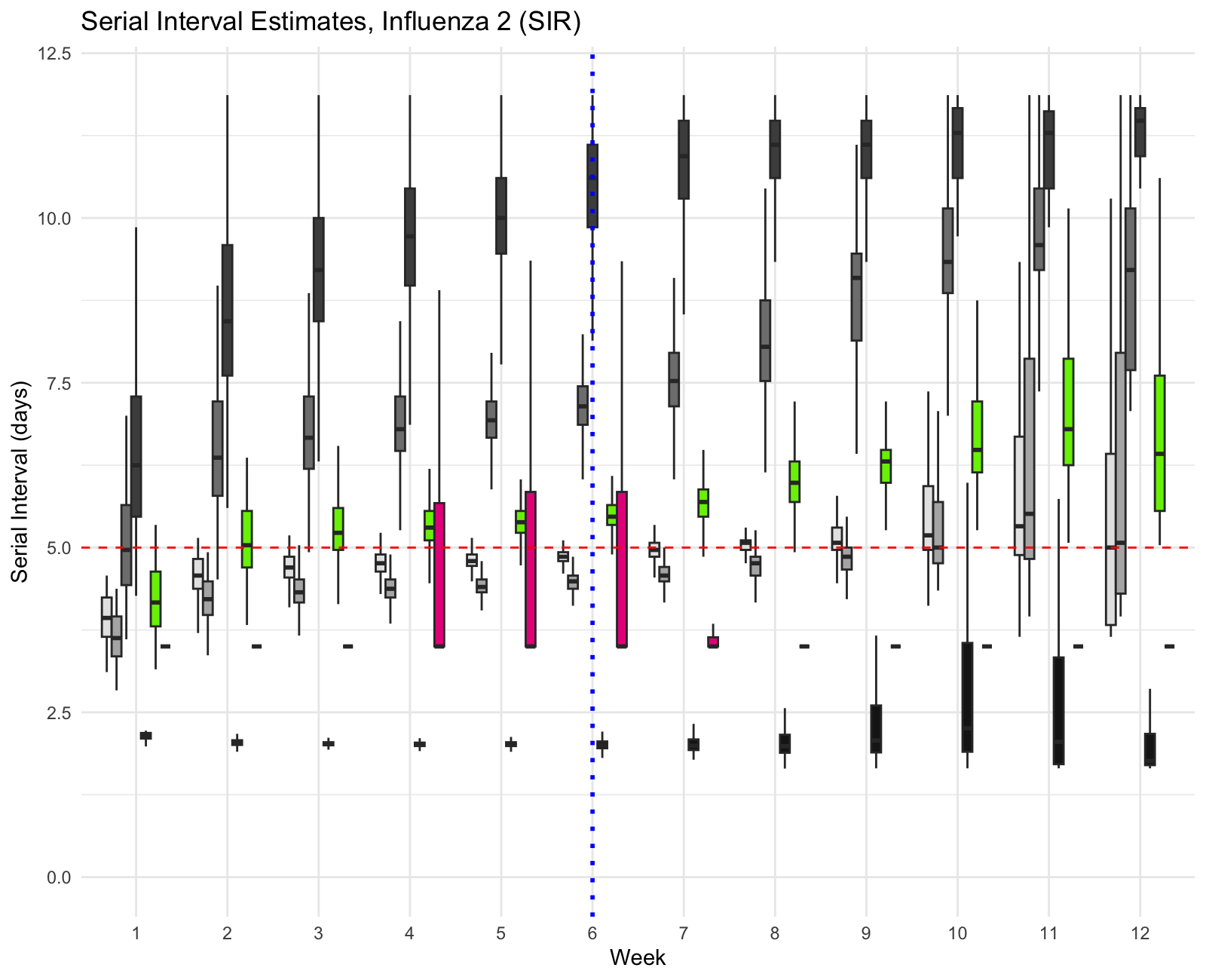}%
            
        } \\ 
        \vspace{1em} 
        \subfloat[Influenza 1, SEIR]{%
            \includegraphics[width=.4\linewidth]{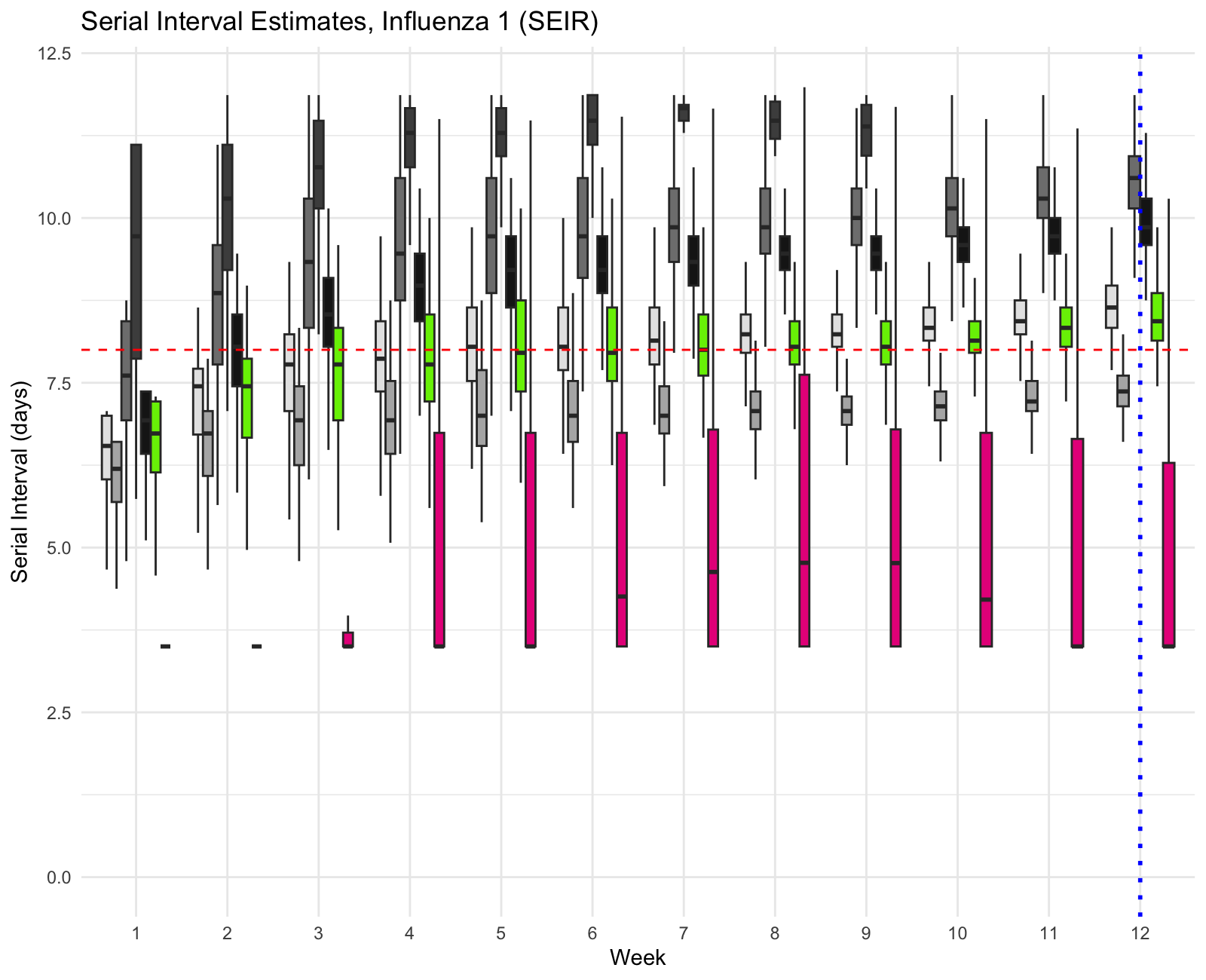}%
            
        }
        \subfloat[Influenza 2, SEIR]{%
            \includegraphics[width=.4\linewidth]{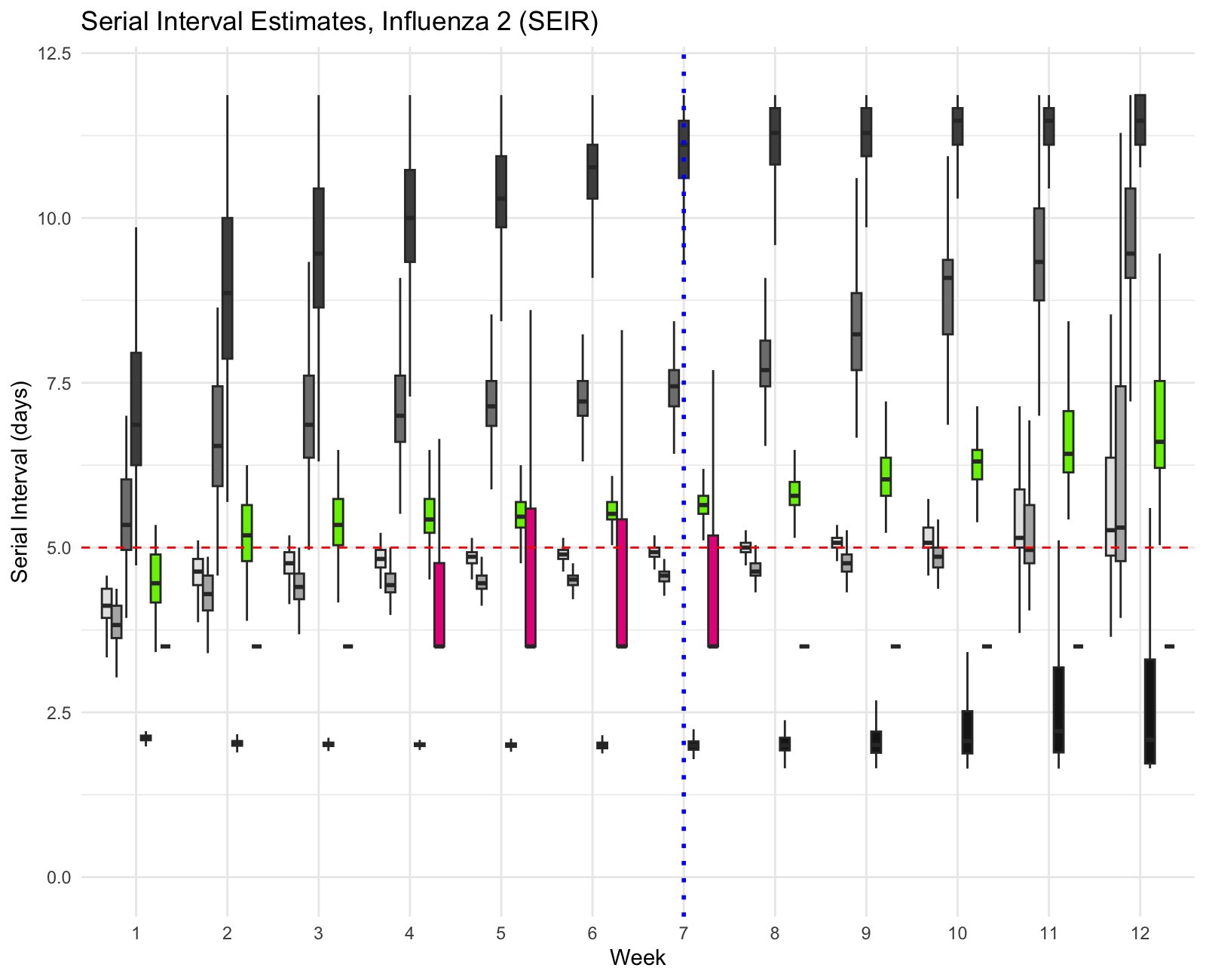}%
            
        } \\ 
        \vspace{1em} 
        \subfloat[Influenza 1, SEAIR]{%
            \includegraphics[width=.4\linewidth]{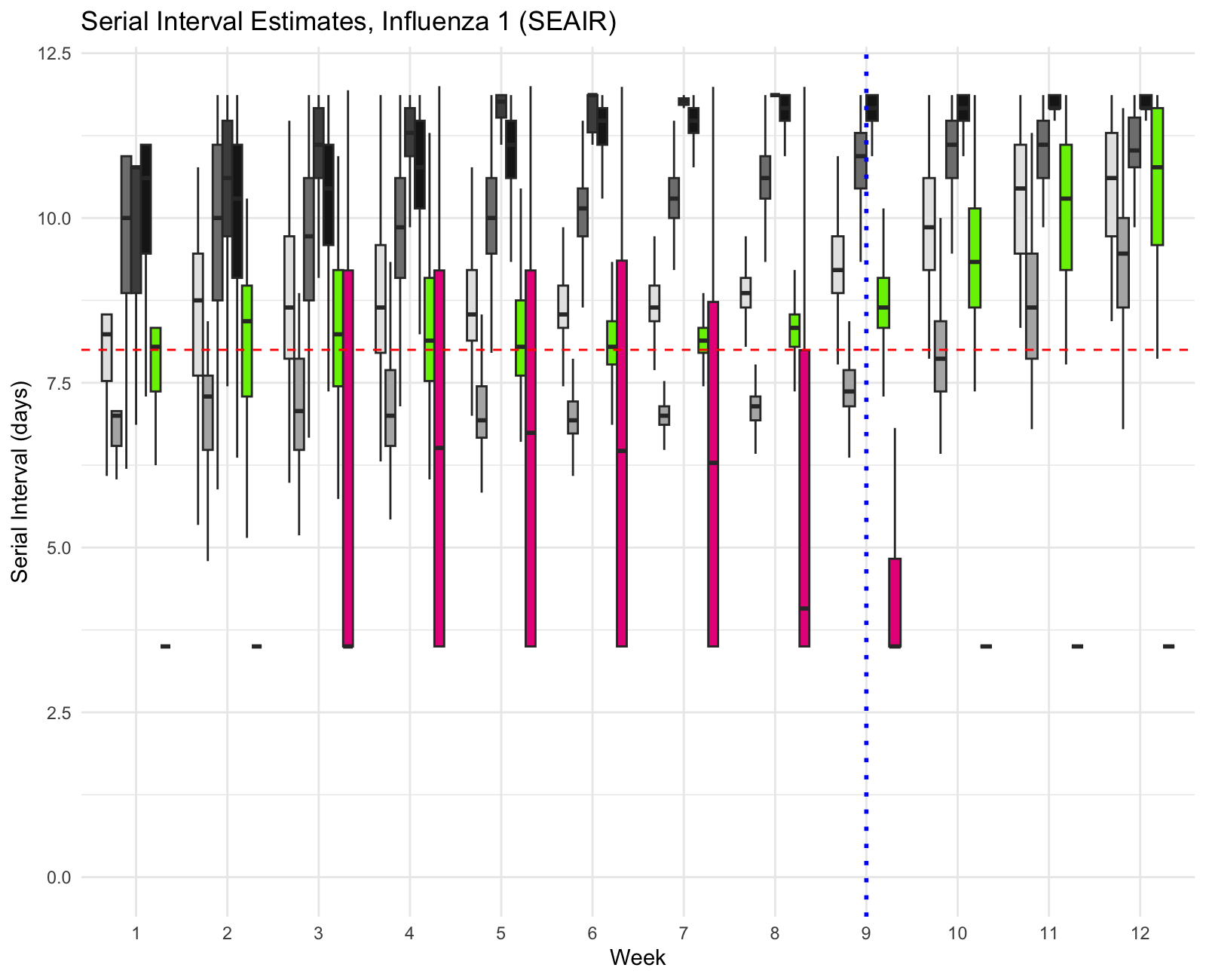}%
            
        } 
        \subfloat[Influenza 2, SEAIR]{%
            \includegraphics[width=.4\linewidth]{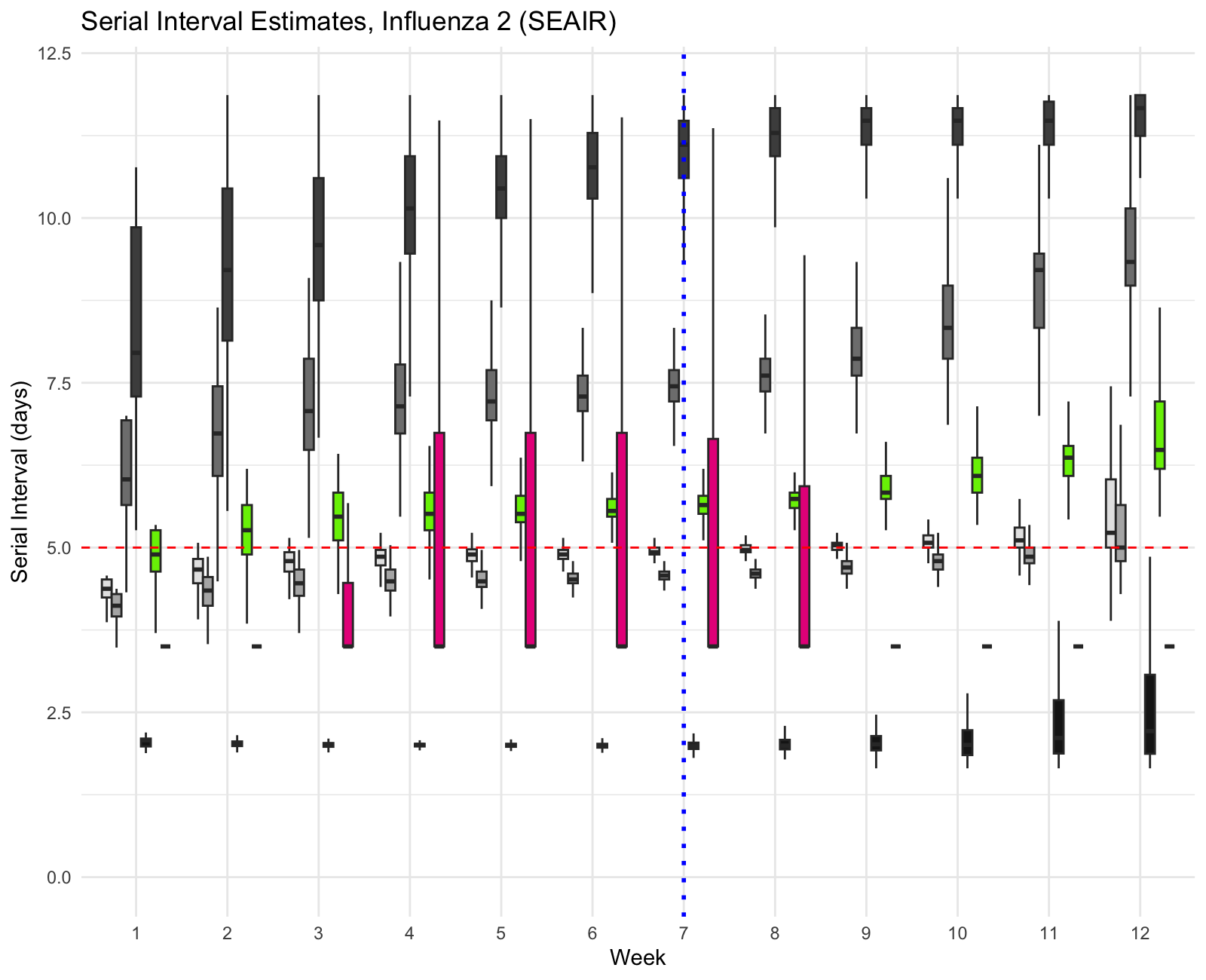}%
            
        } \\
        \vspace{1em}
        \includegraphics[width=0.2\linewidth]{legend.pdf}
        \caption{Comparison of estimates of $SI$ from the SIR, SEIR, and SEAIR datasets of both Influenza 1 and 2. Methods include the White and Pagano method, along with the proposed sequential Bayes method under the varying mispecifications of the prior listed in Table~\ref{tab: misspecifiedCases}.} 
        \label{SI_Influenza-Supp}
    \end{figure}

\end{document}